\documentclass{article}

\def\be{\begin{eqnarray}}
\def\ee{\end{eqnarray}}
\def\nn{\nonumber}

\hoffset= - 1.0in         

\usepackage{ifthen}
\usepackage{graphicx}
\def\DMS#1,#2,#3{\includegraphics[width=#1pt,height=#2pt]}

\newcommand{\Fig}[3]
{\bigskip
\begin{figure}\begin{center}
\ifthenelse{\equal{#2}{}}
{\includegraphics[width=200pt,height=200pt,draft]{./pics/#1.pcx}}
{\DMS#2,0{./pics/#1.pcx}}
\caption{{\footnotesize{#3}}}\label{#1}\end{center}
\end{figure}
\bigskip
}

\title{{\bf Linearized Lorentz-Violating  Gravity
and Discriminant Locus in the Moduli Space of Mass Terms}
\vspace{.5cm}}
\author{{\bf Andrei Mironov}\footnote{ {\small {\it
Lebedev Physics Institute} and {\it ITEP, Moscow, Russia}};
mironov@itep.ru; mironov@lpi.ru}, {\bf Sergey Mironov}\footnote{
{\small {\it Moscow State University} and {\it ITEP, Moscow,
Russia}}; badzilla@rambler.ru}, {\bf Alexei
Morozov}\thanks{{\small {\it ITEP, Moscow, Russia}};
morozov@itep.ru}\ \ and\ {\bf Andrey Morozov}\thanks{{\small {\it
Moscow State University} and {\it ITEP, Moscow, Russia}};
Andrey.Morozov@itep.ru}\phantom{a}  }

\begin{document}

\maketitle

\vspace{-6.5cm}

\begin{center}
\hfill FIAN/TD-24/08\\
\hfill ITEP/TH-68/08\\
\end{center}

\vspace{4.5cm}

\begin{abstract}
\noindent We analyze the pattern of normal modes in linearized
Lorentz-violating massive gravity over the 5-dimensional moduli
space of mass terms. Ghost-free theories arise at bifurcation points
when the ghosts get out of the spectrum of propagating particles due
to vanishing of the coefficient in front of $\omega^2$ in the
propagator. Similarly, the van Dam-Veltman-Zakharov (DVZ)
discontinuities in the Newton law arise at another type of
bifurcations, when the coefficient vanishes in front of $\vec k^2$.
When the Lorentz invariance is broken, these two kinds of
bifurcations get independent and one can easily find a ghost-free
model without the DVZ discontinuity in the moduli space, at least,
in the quadratic (linearized) approximation.
\end{abstract}

\bigskip

\paragraph{1. Introduction.} The theory of massive gravity \cite{PF,MG}
attracts a new attention these days \cite{RT,MGatt} because of the
growing belief in acceleration of Universe expansion (the "dark
energy" phenomenon) \cite{DM}.\footnote{ As we understand, the
argument is as follows. There is an experimental evidence that
cosmological constant is actually non-vanishing. From the point
of view of flat geometry, the cosmological constant makes graviton
massive (in fact it also provides it with a source term, linear in
$h$, therefore a more accurate analysis, including the change of
expansion background, is actually required), then,
Lorentz-invariant massive gravity looks ill, but this can be cured
by switching on especially adjusted Lorenz-violating terms, perhaps
as small as the cosmological constant, what makes their effect small
and consistent with existing observations. Alternatively one could
say that non-vanishing negative cosmological constant implies
that flat background geometry is substituted with AdS one, and
all analysis should be made differently from this new perspective
\cite{AdS}. In fact motivations for the study of
infrared-modified gravity are not exhausted by the dark-energy
problem, for some other examples see \cite{RT} and \cite{othex}. }
However, the violation of general covariance in massive gravity is
long known to produce a number of non-trivial effects like
occurrence of ghosts and the lack of perturbative regime at small
distances \cite{Va}, moreover, the van Dam-Veltman-Zakharov (DVZ)
discontinuities \cite{DVZ,Ni} and the Boulware-Deser instabilities
\cite{BD} arise whenever one tries to eliminate the ghosts. In fact,
these problems can probably be avoided, if one sacrifices the
Lorentz invariance \cite{LV,RT}, what allows to extend the number of
possible mass terms and go around the most unpleasant singularities
in the moduli space. This was demonstrated at the level of the
linearized gravity with quadratic action\footnote{ Throughout the
paper, our convention for the metric signature is $(-,+,\ldots,+)$.}
$$K_{\mu\nu,\alpha\beta}h^{\mu\nu}h^{\alpha\beta} =$$
\centerline{$ =\left\{\frac{1}{2}\Big( k_\mu k_\alpha\eta_{\beta\nu}
+ k_\mu k_\beta\eta_{\alpha\nu} + k_\nu k_\alpha\eta_{\beta\mu} +
k_\nu k_\beta\eta_{\alpha\mu}\Big) - \Big(k_\mu
k_\nu\eta_{\alpha\beta} + k_\alpha k_\beta \eta_{\mu\nu}\Big) -
\frac{1}{2}k^2\Big(\eta_{\mu\alpha}\eta_{\nu\beta} +
\eta_{\nu\alpha}\eta_{\mu\beta}\Big)
+k^2\eta_{\mu\nu}\eta_{\alpha\beta}\right\}h^{\mu\nu}h^{\alpha\beta}
+ $} \be + m_0^2h_{00}^2 + 2m_1^2h_{0i}^2 - m_2^2h_{ij}^2 + m_3^2
h_{ii}^2 - 2m_4^2h_{00}h_{ii}, \label{lingra} \ee where the first
line is nothing but quadratic approximation to the Einstein-Hilbert
action, while the second line contains five different mass
terms,\footnote{ Of course, one can violate Lorentz invariance not
only in the sector of masses, but also in kinetic term and add
higher derivatives in space directions, which do not produce new
ghosts. For profound example of this kind see \cite{Hor}. The
methods of the present paper are straightforwardly applicable to
these non-minimal deformations, but on this road the moduli space
${\cal M}$ is in no way restricted and eigenvalue patterns can be
made arbitrarily complicated. } which violate both gauge (general
coordinate) and Lorentz $SO(d-1,1)$ invariance, but preserve space
rotation symmetry $SO(d-1)$. In our notation, $h_{ii}^2 =
\left(\sum_{i=1}^{d-1} h_{ii}\right)^2$, while $h_{ij}^2 =
\sum_{i,j=1}^{d-1} h_{ij}^2$. The theory also has the $P$ and $T$
reflection symmetries, so that all scalar physical quantities depend
on the squares $\omega^2$ and $\vec k^2$ of frequencies and space
momenta. {\bf The Lorentz invariance} is restored if the five mass
parameters can be expressed through only two independent quantities,
$A$ and $B$: \be
m_0^2 = B-A, \nn \\
m_1^2=m_2^2=A, \nn \\
m_3^2=m_4^2=B
\label{LIca}
\ee
and ${\cal K}_{\mu\nu,\alpha\beta}$ in (\ref{lingra}) reduces to
\be
\frac{1}{2}\Big( k_\mu k_\alpha\eta_{\beta\nu} +
k_\mu k_\beta\eta_{\alpha\nu} +
k_\nu k_\alpha\eta_{\beta\mu} +
k_\nu k_\beta\eta_{\alpha\mu}\Big)
- \Big(k_\mu k_\nu\eta_{\alpha\beta} +
k_\alpha k_\beta \eta_{\mu\nu}\Big) - \nn \\
- \frac{1}{2}(k^2 + A)\Big(\eta_{\mu\alpha}\eta_{\nu\beta} +
\eta_{\nu\alpha}\eta_{\mu\beta}\Big) +
(k^2 + B)\eta_{\mu\nu}\eta_{\alpha\beta}
\ee
{\bf The ghost-free Lorentz-invariant} Pauli-Fierz \cite{PF} massive
gravity corresponds to the choice $A=B$: it, however,
suffers from all the above-mentioned problems and
thus looks unviable \cite{RT}.
The Lorentz-violating theories (\ref{lingra})
can be ghost free when either
$m_0=0$ or $m_1=0$, and the second choice is the
current favorite candidate for a phenomenologically
acceptable version of massive gravity \cite{RT}.

Lorentz violation breaks a lot of familiar properties of quantum
field theory models and looks unusual in many respects. It gives
rise to the whole variety of non-trivial quasi-particles which can
be ghosts, superluminals and even not look like particles at all
(either relativistic, or non-relativistic). In \cite{MMMMT}
we provided a systematic analysis of the theory (\ref{lingra})
and carefully reproduced and explained the results of \cite{RT},
also relating them to the obvious self-consistency of Kaluza-Klein
theories, which involve massive gravitons but remain free of any
kind of pathologies. Here we present this analysis in still another,
concise and formal way, omitting a lot of details and physical
motivations included into \cite{MMMMT}. Note that, due to the
different choice of signature, the sings of eigenvalues throughout
the paper are opposite to those in \cite{MMMMT}.

\paragraph{2. The main quantity: the propagator $\Pi (k)$
over the moduli space.}
We remind briefly the standard string-theory approach to
consideration of {\it a family} of physical theories \cite{UFN2},
adapting it to a particular application to linearized gravity,
perhaps, Lorentz-violating.

The physical content of a particular theory (model)
is best expressed in terms of the partition function
\be
Z(J) = \int D\phi\ e^{i(S(\phi) + \int J\phi)}
\label{pf}
\ee
In quadratic approximation, when
\be
S(\phi) = \int d^d k \ \phi(-k)K(k)\phi(k)
\ee
and
\be
\int d^dx\ J\phi = \int d^d k\ J(k)\phi(k),
\ee
this $Z(J)$ is also a quadratic exponential,
\be
Z(J) = \exp \left( -{i\over 4}\int d^d k \ J(-k) K^{-1}(k) J(k)\right)
\ee
made from  inverse of kinetic matrix $K(k)$, i.e. the propagator.
This is a finite-dimensional matrix in the space of fields $\phi(x)$:
if $\phi^a(x)$ carries an index $a$, then $K_{ab}(k)$ carries two
indices $a,b$. In the case of vector fields $a$ is just the Lorentz
index $\mu$, in the case of gravitational field $a=(\mu\nu)$ is a
symmetric pair of the Lorentz indices, thus taking $\frac{d(d+1)}{2}$
different values, what reduces to
$\frac{(d-1)(d+2)}{2}$ in the case of the traceless field.
Our main task is to investigate the quantity
$\Pi(k) = J(-k) K^{-1}(k) J(k)$.

Of most interest for us in this paper are two kind of characteristics
of $\Pi$.

(i) A singularity of $\Pi(k)$ defines
a propagating particle and the position of the singularity defines
its dispersion relation $\omega = \varepsilon(|\vec k|)$.

(ii) The quantity $V(|\vec k|)=\Pi(\omega=0,\vec k)$ defines
an instantaneous Newton/Coulomb/Yukawa-like interaction.

The partition function $Z$ and its quadratic approximation
$\ \exp(-{i\over 4}\int \Pi)\ $ are
of course defined over the space of theories ${\cal M}$
(and are, hence, generalized $\tau$-functions \cite{gentau},
ordinary and quasiclassical respectively),
and we are going to study the singularities
(reshufflings or bifurcations) of dispersion relations and
potentials $V$ over the moduli space ${\cal M}$.
In the current problem coordinates in ${\cal M}$ parameterize
the kinetic matrix $K(k)$, actually, the mass terms, and,
as usual in string theory, in the spirit of third-quantization,
they can be considered as vacuum averages of some other fields
(slow variables or moduli).

\paragraph{3. The notion of eigenvalues and its ambiguity.}
The problem of dispersion relations is basically that
of the eigenvalues of $K(k)$: roughly, $\omega = \varepsilon(|\vec
k|)$ is a condition that some eigenvalue $\lambda(k) = 0$. However,
this "obvious" statement requires a more accurate formulation. The
point is that $K$ is actually a quadratic form, not an operator,
what means that it can always be brought to the canonical form with
only $\pm 1$ and $0$ at diagonal, thus leaving no room to quantities
like $\lambda(k)$. Still, this "equally obvious" counter-statement
is also partly misleading, because we are interested not in an
isolated quadratic form, but in a family of those, defined over
${\cal M}$. This means that the sets of $\pm 1$ and $0$ can change
as we move along ${\cal M}$, and degeneracy degree of quadratic form
$K(k)$ can change. Of course, this degree (a number of $0$'s at
diagonal) is an integer and changes abruptly -- and thus is not a
very nice quantity. A desire to make it smooth brings us back a
concept of $\lambda(k)$. However, in order to introduce $\lambda(k)$
one needs an additional structure, for example, a metric in the
space of fields.

In application to our needs one can introduce "eigenvalues"
$\lambda(k)$ as follows: consider instead of
$\Pi = J\frac{1}{K}J$ a more general quantity
\be\label{I}
\Pi(\lambda|k) = J\frac{1}{K-\lambda I}J
\ee
Then \textit{as a function of $\lambda$} it can be represented as a sum
of contributions of different poles:
\be
\Pi(\lambda|k) = \sum_{a,b,c}
\frac{\alpha_a^{bc} J_bJ_c}{\lambda_a-\lambda}
\ee
then $\lambda_a(k)$ are exactly the "eigenvalues"
that we are interested in, and our original
\be
\Pi(k) = \sum_{a,b,c}
\frac{\alpha_a^{bc}(k) J_b(-k)J_c(k)}{\lambda_a(k)}
\ee
The only thing that one should keep in mind is that this
decomposition depends on the choice of additional matrix
(metric) $I$, which can be chosen in different ways, in particular,
its normalization can in principle depend on the point of ${\cal M}$.
We shall actually assume that it does {\it not}, and clearly
the physical properties do not depend on this choice, however,
concrete expressions for $\lambda_a(k)$ do. It is important,
that the dispersion relations -- the zeroes of $\lambda_a(k)$ --
are independent of $I$.

Introduction of $I$ is also important from another point of view.
To be well-defined, the Lorentzian partition function
requires a distinction between the retarded and advanced correlators
(Green functions), which is usually introduced by adding
an infinitesimal imaginary term to the kinetic matrix $K$:
the celebrated $i\epsilon$ in the Feynman propagator.\footnote{
The Feynman propagator implies that {\it particles} with the dispersion
relation $\omega = +\varepsilon(|\vec k|)$ propagate {\it forward}
in time, while {\it antiparticles} with
$\omega = -\varepsilon(|\vec k|)$ -- {\it backwards} in time.
Since
$\theta(\pm t) = \frac{1}{2\pi i}\int \frac{e^{i\omega t}d\omega}
{\pm\omega - i \epsilon}$,
we have for the propagator
$$\frac{1}{2\varepsilon}\left(
\frac{1}{\omega - \varepsilon - i\epsilon}
+ \frac{1}{-\omega - \varepsilon - i\epsilon}\right) =
\frac{1}{\omega^2 - \varepsilon^2 - i\epsilon}
$$
For ghosts with the propagator $\frac{1}{\omega^2 - \varepsilon^2 +
i\epsilon}$ the situation is inverse: particles propagate backwards
while antiparticles forward in time. See also Appendix II. }
However, in the case of kinetic \textit{matrix} this is not just $i\epsilon$,
it is rather $i\epsilon I_F$ with some particular matrix $I_F$. If
we identify our $I$ with $I_F$, then the dispersion relations are
actually
\be
\lambda_a(k) = i\epsilon
\label{iep}
\ee
what implies that
$\lambda_a(k)$ is, in fact, very different from $-\lambda_a(k)$, and
this is related to the important concept of {\it ghosts}.

The most natural choices of the matrix $I$ are probably either just the unit matrix,
or "the Lorentzian unit matrix", i.e. that with -1 corresponding to the $0$-components. The physically
justified choice is the unit (Euclidean) matrix, while technically it is often simpler to work with the
Lorentzian unit matrix, especially when dealing with theories with the Lorentz invariance unbroken.
At the same time, in these two cases it is only ghost content of the non-scalar sectors which differs.
Therefore, it is often safe (and technically preferable) to use the Lorentzian unit matrix. We illustrate
this in the simplest warm-up example of the massive vector field theory in Appendix I, where we compare the results
obtained for the two cases of Euclidean and Lorentzian eigenvalues (Euclidean and Lorentzian unit matrices).
Since this paper is rather
devoted to the method than to concrete physical applications, we use the Lorentzian eigenvalues here, leaving
the Euclidean ones for \cite{MMMMT}, where we deal with physical issues.

\paragraph{4. Spectrum and the phase diagram.}
Important information about the
theory is contained in its {\it spectrum}: positions of the poles of
$\Pi(\lambda|k)$ in the complex $\lambda$-plane. These positions
define the dispersion relations $\lambda_a(\omega,\vec k)=0$ between
the frequency $\omega$ and the wave vector $\vec k$ of elementary
excitations (quasiparticles) and the way these relations depend on
the point of the moduli space ${\cal M}$.

As one knows well from condensed matter physics, in generic
Lorentz-violating theory dispersion relations are quite
sophisticated, they are roots of polynomial equation and often do
not possess any useful analytical expressions. Sometime they are
better represented by pictures: the plots $\lambda_a(\omega)$ or
$\lambda_a(\vec k)$, however when the pattern is multi-dimensional
one can only draw its particular $2d$ or $3d$ sections, which do not
provide complete visualization. Developed algebra-geometric
intuition is actually needed to analyze the spectrum -- surprisingly
enough, this is already the case in such a fundamental (and seemingly
simple) theory as linearized gravity!

Of main interest are {\it qualitative} features of the spectrum and
their {\it bifurcations}: the changes of these qualitative features
when one goes from one region of the moduli space to another. The
corresponding division of the moduli space into domains with
qualitatively different spectra (and, perhaps, other physically
relevant characteristics like structure functions
$\alpha_a^{bc}(k)$) is called the {\it phase diagram} of the theory
(or, better, of the family of theories).

\paragraph{5. Ghosts, tachyons, superluminals and DVZ jumps \label{ghosts}}

The simplest examples of qualitative features of the spectrum are
the presence or absence of exotic (from the perspective of
Lorentz-invariant field theory) excitations, like ghosts or
superluminals.

The ghost differs from the normal particle by a sign in front of
$\omega^2$ in $\lambda_a(k)$. For example, for a scalar particle,
\be
\left.\frac{\partial\lambda_a(k)}{\partial\omega^2}
\right|_{\lambda_a\!(k)=0} \ \ \ \ \ \
\begin{array}{ccc}
< 0 & & {\rm normal\ particle} \\
> 0 & & {\rm ghost}
\end{array}
\label{ghoco}
\ee
In order to define this sign one needs to compare
it with the one in front of $i\epsilon$ in (\ref{iep}), see Appendix II
for a more detailed discussion. Problems are actually expected
when excitations with opposite signs are present: when the "ghosts"
need to coexist and {\it interact} with the "normal" particles. The
condition
\be
\left.\frac{\partial\lambda_a(k)}{\partial\omega^2}
\right|_{\lambda_a\!(k)=0} = 0
\label{dldo}
\ee
defines the loci in
the moduli space, where the ghost content of theory can change.

A similarly-looking condition
\be
\left.\frac{\partial\lambda_a(\omega=0,\vec k)} {\partial\vec k^2}
\right|_{\lambda_a\!(k)=0} = 0
\label{dldk}
\ee
defines the loci of the {\bf DVZ jumps}, see below.
In the Lorentz invariant theory, where $\lambda_a$ depends on $k^2 =
-\omega^2+\vec k^2$, the two conditions (\ref{ghoco}) and
(\ref{dldk}) are clearly related. Therefore, one can easily come
across the DVZ jump when trying to get rid of ghosts -- and this,
indeed, happens in the simplest Pauli-Fierz version of linearized
gravity. After the Lorentz violation, the link between (\ref{ghoco})
and (\ref{dldk}) is relaxed.

Next, the difference between the normal particles and tachyons is as
follows:
\be
\begin{tabular}{c}
if $\lambda_a(\omega, \vec k=0) = 0$ has real solutions for
the frequency $\omega$, this is a normal particle,\\
if $\lambda_a(\omega=0, \vec k) = 0$ has real solutions for the wave
vector $\vec k$, this is a tachyon.
\end{tabular}
\ee

The {\bf superluminal propagation} \cite{suli} is controlled by the
group velocity \be \vec v_a = \left.\frac
{{\partial\lambda_a(k)}/{\partial\vec k}}
{{\partial\lambda_a(k)}/{\partial\omega}} \right|_{\lambda_a\!(k)=0}
\ee in the usual way: \be \vec v_a^2 \ \ \ \ \ \begin{array}{ccc}
<1 & & {\rm normal\ particle} \\
=1 & & {\rm light-like\ particle} \\
>1 & & {\rm superluminal\ particle}
\end{array}
\ee In fact, it makes sense to further distinguish between different
superluminals by looking at another quantity: \be V_a^2 =
\left.\frac {{\partial\lambda_a(k)}/{\partial\vec k^2}}
{{\partial\lambda_a(k)}/{\partial\omega^2}}
\right|_{\lambda_a\!(k)=0} \ee For the ordinary relativistic
particle with $\lambda = -\omega^2+\vec k^2+m^2$, this $V^2=1$
independently of the value and even of the sign of mass $m^2$. Thus,
the ordinary tachyons with negative $m^2$ and $\vec v^2>1$ are
rather "soft" superluminals. In the Lorentz violating theories
things are much worse: there are "harder" superluminals with
$V^2>1$.

Finally, the DVZ jump can occur when one of the scalars becomes
infinitely heavy. Then the massless limit, when all the five moduli
$m_0,\ldots,m_4 \rightarrow 0$, gets ambiguous: this scalar can
either remain infinitely heavy or acquire a finite mass or become
massless, depending on a particular way the limit is taken. Thus,
the contribution of such a scalar to the instantaneous potential is
also ambiguous and depends on the way one approaches the point
$m_0,\ldots,m_4=0$: if we are interested in physically relevant
quantities, this point in ${\cal M}$ is, in fact, singular and
should be blown up to resolve the singularity. For the generic
dispersion relation the role of mass in above reasoning is played by
the root $\vec k_0^2$ of the equation $\lambda_a(\omega=0,\vec
k_0^2)= 0$ (the real mass gap arises when the root is negative,
$\vec k_0^2<0$). The DVZ jump can occur when $\vec k_0^2\rightarrow
-\infty$, and this actually requires that $\lambda_a(\omega=0,\vec
k^2)$ has an asymptote which satisfies (\ref{dldk}).
Thus, (\ref{dldk}) is a necessary condition for a
the DVZ jump to occur. Note, however, that (\ref{dldk}) is more
restrictive, because $\omega=0$ condition is additionally imposed:
thus it defines a codimension-one subspace in the moduli space
${\cal M}$, while (\ref{ghoco}) can hold for particular $\omega$ and
$\vec k$ in codimension-zero domains of ${\cal M}$. The DVZ jump is
basically a non-commutativity of the limits,
i.e. the difference between the two naive definitions of the static
potential (the instantaneous Newton/Coulomb/Yukawa interactions) at
a given point $M_0$ in the moduli space. Such a difference can occur
when the number of degrees of freedom changes at $M_0$, i.e. when
the two branches of dispersion relations merge or intersect. This
happens if the two roots $\lambda_a(\omega=0,\vec k)$ coincide, i.e.
when (\ref{dldk}) takes place.

In Appendix I we thoroughly study conditions of emergency of ghosts, tachyons,
superluminals and the DVZ jumps in the example of massive vector theory.

\paragraph{6. Eigenvalues and discriminant analysis.} The "eigenvalues"
$\lambda_a(k)$ are roots of the characteristic polynomial
\be
C_I(\lambda) = {\rm discriminant}_\phi
\Big(S(\phi) - \lambda(\phi I\phi)\Big) = \det (K-I\lambda)
= \prod_a^{{\rm deg}\ C} (\lambda - \lambda_a),
\ee
since the
discriminant of a quadratic form is actually a determinant of the
corresponding matrix. On-shell conditions $\lambda_a(k)=0$ are
zeroes of ${\rm discriminant}_\phi\Big(S(\phi)\Big) = \det K$
itself and do not depend
on the choice of $I$, as we already mentioned.

Similarly, conditions like (\ref{dldo}) and (\ref{dldk}) are zeroes
of the ratio \be \frac{{\rm resultant}_\lambda
\Big(C(\lambda),\delta C(\lambda)\Big)} {{\rm
resultant}_\lambda\Big(C(\lambda),C'(\lambda)\Big)}\ = (-)^{1+{\rm
deg}\ C} \prod_a \delta\lambda_a \label{resra} \ee where $\delta$ is
any variation of the coefficients of $C(\lambda)$, say, resulting
from an infinitesimal change of $\omega^2$ or $\vec k^2$, and
$C'(\lambda)$ is $\lambda$-derivative of $C(\lambda)$. Note that the
resultant in the numerator has degree ${\rm deg}(C)= \#(a)$ in the
coefficients of $\delta C$, and the resultant in denominator is
actually a discriminant of $C(\lambda)$. For definitions of
resultants and discriminants see, e.g., \cite{discres,DoM}.

\paragraph{7. The pattern of eigenmodes for Lorentz-violating gravity.}
After these general remarks, we return to the concrete model: the
linearized massive gravity (\ref{lingra}).

Eigenvectors of the kinetic matrix are naturally split into three
groups: traceless tensors, vectors and scalars.
In more detail, the $\frac{d(d+1)}{2}$ components of symmetric
tensor $h_{\mu\nu}$ are decomposed as follows:
\be
\frac{d(d+1)}{2}\ \ \  = \ \ \
\underbrace{\frac{(d-2)(d+1)}{2}}_{{\rm massive\
spin}\ 2}
 + \underbrace{\underbrace{(d-1)}_{{\rm space-time\ transverse}}
 +\underbrace{1}_{{\rm secondary}} }_{{\rm
 Stueckelberg\ vector}} +
\underbrace{1}_{{\rm  space-time\ trace}} =
\label{degf0}
\ee
$$ =
\left\{
\underbrace{\frac{d(d-3)}{2}}_{{\rm spatial-transverse\ tensor}} +
\underbrace{(d-2)}_{{\rm longitudinal\ tensor}\atop{={\rm trasverse\ vector}}} +
\underbrace{1}_{{\rm spatial\ trace}}\right\} +
\left\{\underbrace{d-2}_{{{\rm spatial-transverse}}\atop
{{\rm Stueckelberg\ vector}}}
+ \underbrace{1}_{{\rm longitudinal}\atop{\rm Stueckelberg\ scalar}}
+ \underbrace{1}_{{\rm secondary}\atop{\rm Stueckelberg\ scalar}}\right\} + 1
$$
where the first line is the $SO(d-1)$ classification in the rest frame (where $\vec k=0$),
while the second line is the classification in the arbitrary frame, i.e. that w.r.t.
$SO(d-2)$, which acts in the hyperplane transverse to $\vec k$.
Accordingly the characteristic polynomial $C(\lambda)$ in the generic frame is
decomposed as
\be
C(\lambda) =
(\lambda-\lambda_{gr})^{\frac{d(d-3)}{2}}
P_2(\lambda)^{d-2}Q_4(\lambda) =
(\lambda-\lambda_{gr})^{\frac{d(d-3)}{2}}
(\lambda - \lambda_{vec}^+)^{d-2}(\lambda-\lambda_{vec}^-)^{d-2}
\prod_{a=1}^4(\lambda - \lambda_{sc}^a)
\ee
where $P_2$ and $Q_4$ are polynomials of degree $2$ and $4$
respectively and all their coefficients as well as $\lambda_{gr}$
are quadratic functions of $\omega$ and $\vec k$.
Since the coefficients of $C$ are quadratic functions of $\omega$ and $\vec k$,
this means that

$\lambda_{gr}$ is some bilinear combination of $\omega$ and $\vec k$,

$\lambda_{vec}^{\pm} = p_2 \pm \sqrt{p_4}$, where $p_2$ and $p_4$
are respectively quadratic and quartic in $\omega$ and $\vec k$,

$\lambda_{sc}^{1,2,3,4}$ are the roots of degree-four polynomial.

\bigskip

In the rest frame the roots should be grouped in a different way,
according to the first line in (\ref{degf0})
\be
C_{RF}(\lambda) = \left.(\lambda-\lambda_{gr})^{\frac{(d-2)(d+1)}{2}}
(\lambda - \lambda_{vec})^{d-1}(\lambda-\lambda_{sc}^+)
(\lambda-\lambda_{sc}^-)\right|_{\vec k = 0} = 0
\ee
i.e. at $\vec k = 0$
\be
\lambda_{vec}^+(\vec k=0) = \lambda_{gr}(\vec k=0), \nn \\
\lambda_{sc}^{spT}(\vec k=0) = \lambda_{gr}(\vec k=0), \nn \\
\lambda_{sc}^{spS}(\vec k = 0)=\lambda_{vec}^-(\vec k = 0)
\ee
where "spT" and "spS" label the spatial trace $h_{ii}$ and
the spatial Stueckelberg scalar $h_{0i} = k_is$ respectively.
The remaining two scalars, the space-time trace (stT) $h^\mu_\mu$
and the secondary Stueckelberg scalar (seS)
$h_{\mu\nu} = k_\mu k_\nu\sigma$
have eigenvalues, which are roots of quadratic equation:
\be
\left.\lambda^\pm_{sc} = q_2 \pm \sqrt{q_4}\right|_{\vec k = 0}
\label{rfqq}
\ee
In Lorentz invariant theory one can obtain eigenvectors and
eigenvalues in an arbitrary frame by a Lorentz boost, but Lorentz violation forbids
such a simple procedure.

\bigskip

In gauge invariant theory all the $d$ Stueckelberg fields have vanishing
eigenvalues and one gets
\be
C_{GI}(\lambda) = \lambda^d
(\lambda-\lambda_{gr})^{\frac{d(d-3)}{2}}
(\lambda - \lambda_{vec}^+)^{d-2} (\lambda-\lambda_{sc}^{spT})
(\lambda-\lambda_{sc}^{stT})
\ee
i.e. in this case the two trace (spT and spS) eigenvalues are the roots
of quadratic equation,
\be
\lambda_{sc}^{\pm T} = \left.t_2 \pm \sqrt{t_4}\right|_{GI}
\label{GItt}
\ee
Actually gauge invariant will be only the massless
gravity (where, by the way, transition to the rest frame
is not a justified operation). This can be summarized in the following scheme:

\bigskip

\centerline{
\begin{tabular}{|ccccc|}
\hline &&&&\\
rest frame&&normal modes&&gauge invariant (massless) case \\
&&&&\\
&&graviton&$\rightarrow$& graviton\\
&$\swarrow$&&&\\
massive graviton &$\leftarrow\oplus$&&&\\
&$\nwarrow$&&&\\
&&vector&$\rightarrow$&vector\\
&&&&\\
&$\swarrow$&Stueckelberg vector&&\\
Stueckelberg $(d-1)$-vector&&&$\searrow$&\\
&$\nwarrow$&Stueckelberg scalar&$\rightarrow$&Stueckelberg $d$-vector\\
&&&$\nearrow$&\\
secondary Stueckelberg scalar&$\leftarrow$&secondary Stueckelberg scalar&&\\
&&&&\\
&$\oplus\leftarrow$&spatial trace&$\rightarrow$&spatial trace\\
&&&&\\
space-time trace&$\leftarrow$&space-time trace
&$\rightarrow$&space-time trace\\
&&&&\\
\hline
\end{tabular}
}

\bigskip

\begin{figure}\begin{center}
{\includegraphics*[width=0.3\textwidth]{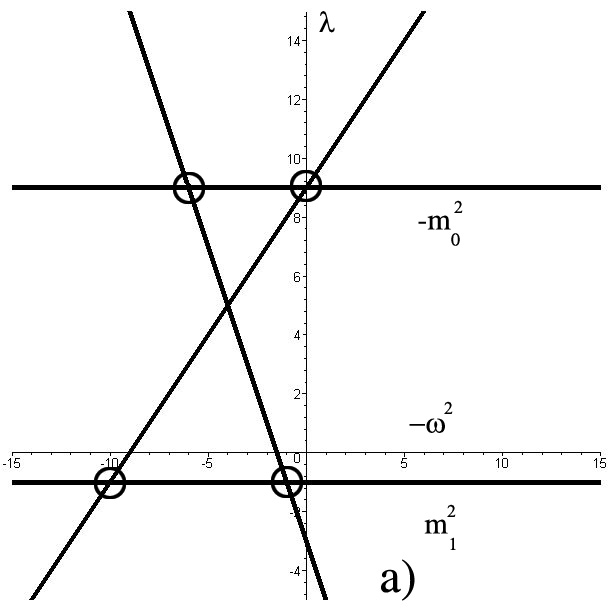}}
\hspace{0.5cm}
{\includegraphics*[width=0.3\textwidth]{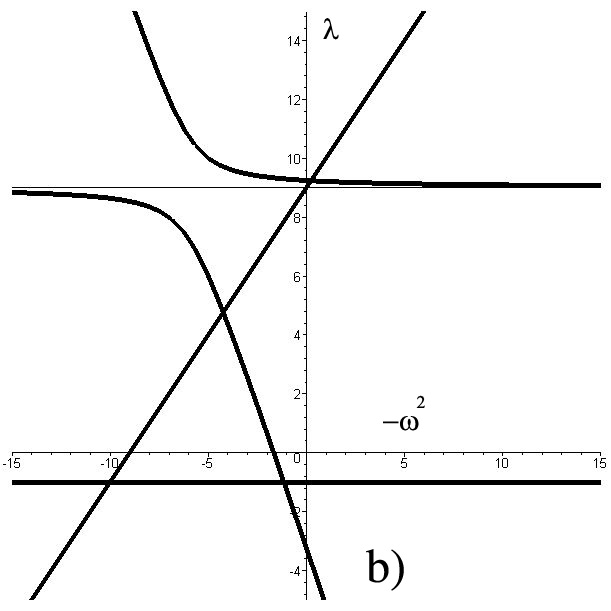}}
\hspace{0.5cm}
{\includegraphics*[width=0.3\textwidth]{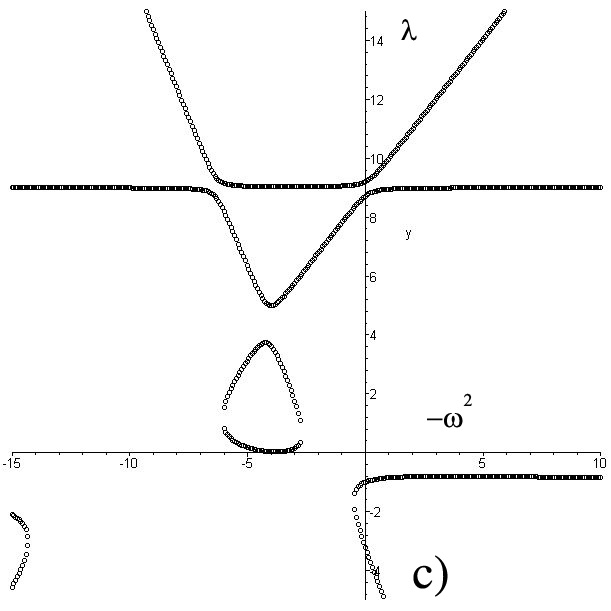}}
\caption{\footnotesize{The left picture plots the four eigenvalues
as functions of $-\omega^2$ in the Lorentz-non-invariant case in the
rest frame and with $m_4=0$. In this case, the figure is maximally
degenerated, and all the eigenvalues are straight lines. The pattern
is described by positions of the two horizontal lines given by
values of $m_0^2$ and $m_1^2$, and by positions of the four
intersections which depend also on $m_2^2$ and $m_3^2$. In the middle figure,
the degeneration is partly lifted by choosing non-zero $m_4$ (still in the rest
frame). At last, in the right
figure, a typical perturbation of the previous figures shown, when
both non-zero $m_4^2$ and momentum are switched on resolving all the four marked
crossings of the left figure. The parameters here are: $m_0^2=-9$, $m_1^2=-1$,
$m_2^2=9$, $m_3^2=4$.}}
\label{nonperturbed}
\end{center}\end{figure}

The most interesting sector is that of scalars, with complicated
inter-mixture of four eigenvectors.
The pattern of eigenvalues is most simple at
$m_4=0$ and $\vec k=0$:
\be
\lambda_{sT} = -\omega^2 + m_2^2, \nn \\
\lambda_{S} = m_1^2, \nn \\
\lambda_{sS} = -m_0^2, \nn \\
\lambda_{stT} = (d-2)\omega^2 + m_2^2-(d-1)m_3^2 \ee see
Fig.\ref{nonperturbed}a.

Switching on $m_4$ and $\vec k$ leads to a bifurcation: repulsion of
levels, so that Fig.\ref{nonperturbed}a is immediately transformed into
Fig.\ref{nonperturbed}b. What is important, however, the horizontal
asymptotes stay at their positions: at $\lambda=-m_0^2$ and $\lambda
= m_1^2$.

Bifurcations become visible in the physical spectrum
when one of these asymptotes
coincide with the real axis, $\lambda = 0$.
Clearly, this happens when either $m_0=0$ or $m_1=0$.

In this paper we introduce eigenvalues in a Lorentz-invariant way,
taking $I = I_L = \eta = {\rm diag}(-1,1,\ldots,1)$, even despite
the Lorentz symmetry can be explicitly violated by mass terms in the
Lagrangian. This should be kept in mind in comparison to
\cite{MMMMT}, where "Euclidean" eigenvalues were considered,
associated with the choice $I = I_E = {\rm diag}(1,1,\ldots,1)$ (see also Appendix II).

\begin{figure}\begin{center}
\hspace{-1cm}
{\includegraphics*[width=0.3\textwidth]{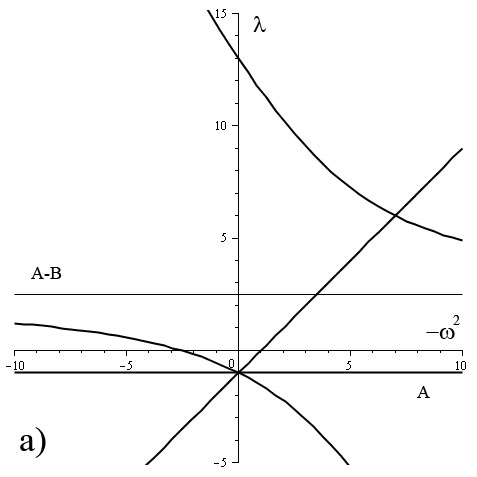}}
\hspace{1cm}
{\includegraphics*[width=0.3\textwidth]{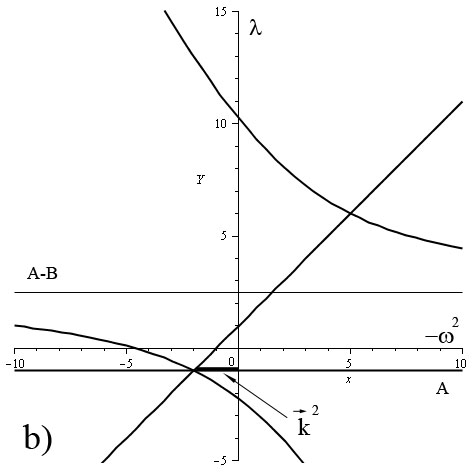}}
\hspace{1cm}
{\includegraphics*[width=0.3\textwidth]{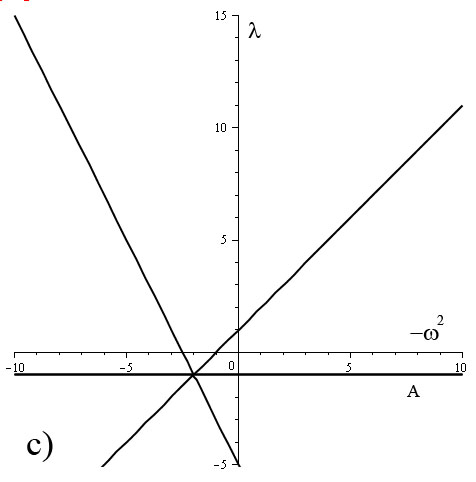}}
\caption{\footnotesize{The left figure plots the four eigenvalues in
the Lorentz-invariant case at zero momentum, and the middle figure
at non-zero momentum (which corresponds just to shifting the whole
figure to the left). The right figure corresponds to the degenerated
case of $B=0$. {\bf Note that in this case there is no way to
resolve all the intersections of the right picture.}}}
\label{LIscals}
\end{center}\end{figure}

\paragraph{8. Restriction to subspace of Lorentz invariant theories in ${\cal M}$.}
To reveal the physical meaning of pictures like Fig.\ref{nonperturbed}, it is
instructive to begin with the simpler
Lorentz-invariant case (\ref{LIca}) with only two moduli
$A$ and $B$.
In this model the Lorentz symmetry expresses eigenvalues
in arbitrary frame through those
in the rest frame, so that the classification of eigenvectors
is always described by the left column of the table.
It remains only to evaluate concrete
functions of $k^2=-\omega^2+\vec k^2$:
\be
\begin{array}{ccc}
{\rm tensors} & (d+1)(d-2)/2 & \lambda_{gr} = k^2 + A \\ &&\\
{\rm Stueckelberg\ vector} & d-1 & \lambda_{vec} = A \\ &&\\
{\rm scalars} & 2 & \lambda_{sc}^\pm = A -\frac{(d-2)k^2 + dB \pm
\sqrt{ (d-2)^2(k^2+B)^2 + 4(d-1)B^2}}{2}
\end{array}
\label{LIei}
\ee
Thus in the Lorentz invariant case we have two scalars: the
space-time trace ("$+$" sign in above formula) and the secondary
Stueckelberg ("$-$" sign). When the Lorentz invariance is violated, we
will get $4$ scalars, two additional coming from the tensor and vector
multiplets (the spatial trace and spatial Stueckelberg scalar
respectively). In Fig.\ref{LIscals} we plot eigenvalues corresponding to these
four scalars as functions of $k^2$ in the Lorentz invariant case (we include those two
scalars that are parts of the tensor and vector multiplets in this case).
The parameter $A$ enters as a common
shift of the horizontal axis, and the mass-shell condition $\lambda = 0$
for propagating particles is satisfied at the intersections of
the eigenvalue curves in the picture with the abscissa axis.
We see that the four lines become straight, like in
Fig.\ref{nonperturbed}a, only at $B=0$, Fig.\ref{LIscals}c:
this is the only point where the condition $m_4=0$ is consistent
with (\ref{LIca}). For finite $B$, the two eigenvalues $\lambda_{sc}$
are repulsed, like in Fig.\ref{nonperturbed}b, while nothing happens
to the other two eigenvalues, protected by the Lorentz symmetry: only
two scalars can mix in this symmetric situation.

We see that of these two scalars only one can be on-shell
and, whatever it is, it is a ghost, see (\ref{ghoco}): the slope of the curve
$\lambda(-\omega^2)$ is negative everywhere and thus also on mass-shell, where
$\lambda(-\omega^2)=0$. It is a tachyon or not,
depending on where the intersection with the abscissa axis
occurs: to the right (tachyon) or to the left (normal)
of the ordinate axis (for positive or negative
$k^2=-\omega^2+\vec k^2$).
The only chance for this ghost to disappear from the spectrum
of propagating particles is when the thin line (asymptotics of the eigenvalues)
coincides with the abscissa axis, i.e. when $A=B$: this is exactly
the Pauli-Fierz model \cite{PF}. Clearly, at this point
(\ref{dldo}) is fulfilled.
However, exactly at the same point in the moduli space
condition (\ref{dldk}) is also satisfied, and the DVZ jump
occurs (it comes with no surprise, because (\ref{dldo}) and (\ref{dldk})
always coincide if the Lorentz invariance is not violated).

\begin{figure}\begin{center}
{\includegraphics[bb= 0 0 10cm 10cm,scale=0.4]
{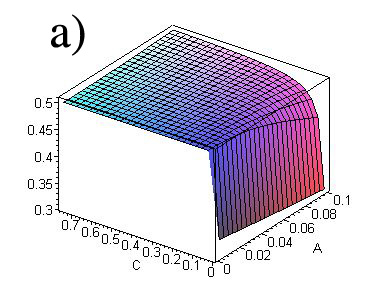}}\hspace{1cm}
{\includegraphics[bb= 0 0 10cm
10cm,scale=0.4] {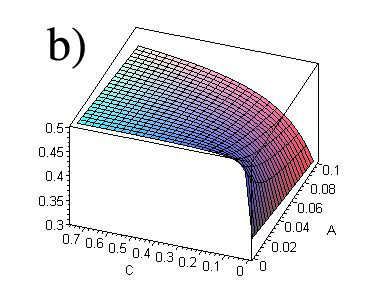}}\hspace{1cm}
{\includegraphics[bb= 0 0 10cm 10cm,scale=0.4]
{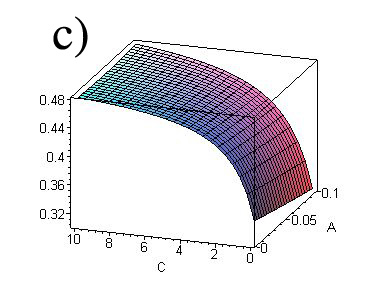}}\hspace{1cm}
\caption{\footnotesize{Resolution of singularity in potential $V$ at
$A=B=0$. Plotted are potentials $V$ at $k^2=1$ as functions of $A$
and $C$, where $C$ substitutes $B$ and is introduced in three
different ways: $B=A+C$ (Fig.{\bf a}), $B=A+AC$ (Fig.{\bf b}), $B=A+A^2C$
(Fig.{\bf c}). Clearly, $V$ is a smooth function at $A=0$ only in the
third case. Of course, $V$ is also singular when propagating
particles contribute, i.e. at $k^2+m^2=0$ and $k^2+M^2=0$. These
singularities is avoided since we present only a fragment of plots
with small enough $A \ll k^2=1$. }} \label{NYpot}
\end{center}\end{figure}

The DVZ jump occurs because the instantaneous-interaction potential
$V(\vec k)=\Pi(\omega=0,\vec k)$ does not have a well defined limit
when both $A$ and $B$ tend to zero. Instead, $V(\vec k)$ is well
defined on a properly compactified moduli space with a blown-up singularity
at $A=B=0$: if
one parameterize $B$ as $B = A + A^2\xi$, then $V$ is actually a
smooth function of $A$ and $\xi$, see Fig.\ref{NYpot}c. In more
detail, the Newton/Yukawa-potential is given by \be V(k)=
\frac{J_0(k)J_0(k)}{(d-1)(d-2)} \left(\frac{d-2}{k^2+m^2} +
\frac{1}{k^2+M^2}\right), \ \ \ \ \ \
m^2 = A, \ \ \ \ M^2 = \frac{A(dB-A)}{(d-2)(A-B)} \label{VvsAB} \ee
see \cite[s.3.6]{MMMMT}. It is plotted as a function of $A$
and $B$ in Fig.\ref{NYpot}a at some fixed value of $k^2$. Poles at
the two lines $k^2+m^2=0$ and $k^2+M^2=0$ correspond to propagating
degrees of freedom, the second singularity may exist even at
$\omega^2=0$ i.e. at positive $k^2$, because the ghost can be also a
tachyon, with $M^2<0$ (it is {\it not} the case if $B\leq A \leq dB$).
Clearly, the function $V$ is discontinuous at $A=B=0$, but the
singularity is resolved in the coordinates $(A,\xi)$ in Fig.\ref{NYpot}c,
at expense of gluing in a whole line $(A=0,\xi)$ instead of a
single point $A=B=0$ (the singularity point is "blown up"). The situation
is of course similar to the resolution of the singularity at $A=B=0$ in the
rational function $\frac{A-B}{A+B}$ by passing, say, to polar or
non-homogeneous coordinates $B=A\xi'$, the difference here is that
the singularity in (\ref{VvsAB}) is rather cusp-like, see
Fig.\ref{NYpot}b, and the blow-up procedure is slightly more involved.

\begin{figure}\begin{center}
\vspace{-2cm}
{\includegraphics*[width=0.3\textwidth]{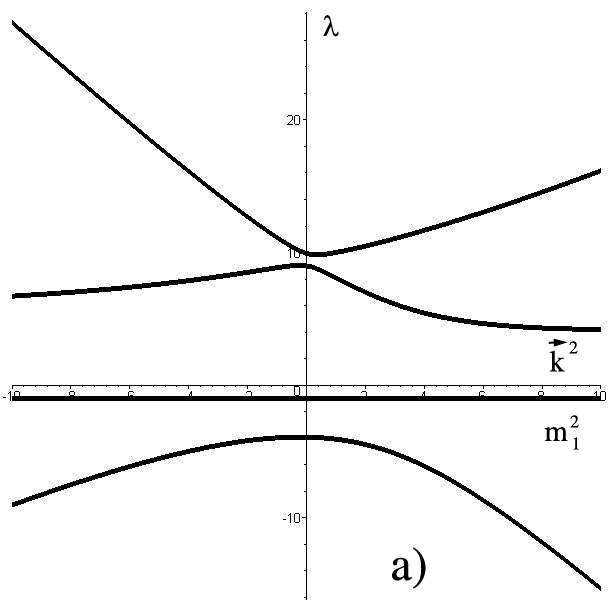}}
{\includegraphics*[width=0.3\textwidth]{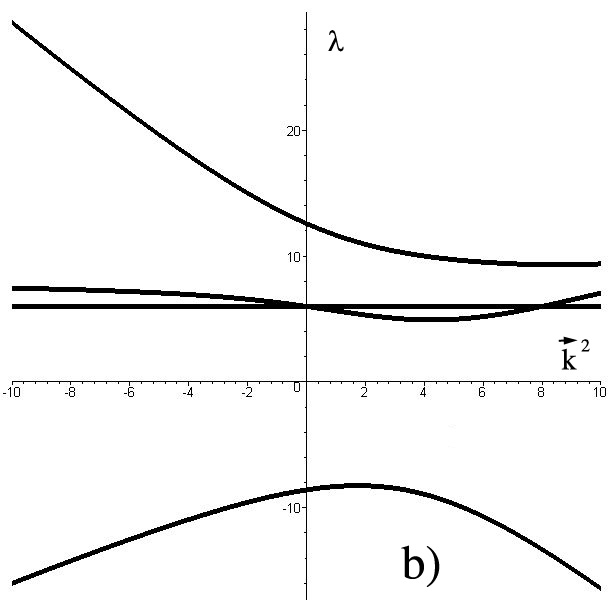}}
\vspace{1cm}
\\
{\includegraphics*[width=0.3\textwidth]{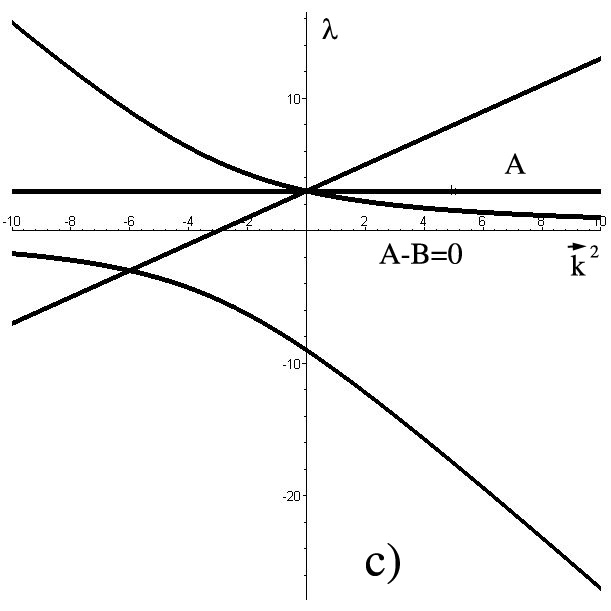}}
{\includegraphics*[width=0.3\textwidth]{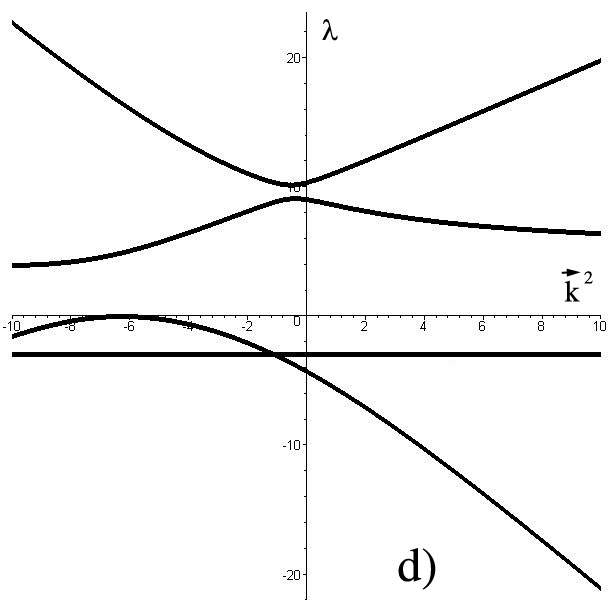}}
\caption{\footnotesize{The eigenvalue curves at $\omega^2=0$ (which are relevant to
describing the potential). The first two figures ({\bf a} and {\bf b})
correspond to the generic Lorentz violating
case (the values of parameters are $m_0^2=-9$, $m_1^2=-1$, $m_2^2=9$, $m_3^2=4$,
$m_4^2=-2$ and $m_0^2=8$, $m_1^2=m_2^2=6$, $m_3^2=m_4^2=-2$.
The third figure ({\bf c}) describes the Lorentz invariant
case, when the eigenvalue asymptotics coincide with the abscissa axis. This corresponds to
the DVZ jump and, at the same time, to the Pauli-Fierz theory ($A=B=3$).
Figure {\bf d} ($m_0^2=-9$, $m_1^2=-3$, $m_2^2=9$, $m_3^2=4$, $m_4^2=2.4$)
demonstrates that condition (\ref{dldk})
can be also realized in a different way (when one of the eigenvalue curve touches the abscissa
axis).
}} \label{kDepend}
\end{center}\end{figure}

\paragraph{9. Back to generic Lorentz violating theory.}
Coming back to the Lorentz-violating masses (\ref{lingra}),
we obtain a somewhat richer pattern of bifurcations, but
their physical interpretations remain very similar.
The essentially new thing is that the singular subspace of the moduli
space has higher codimension and can be passed by in
an easier way.

We begin with the analogue of Fig.\ref{LIscals}a in the rest
frame: see Fig.\ref{nonperturbed}b.
To keep pictures similar to the Lorentz-invariant case, we
plot dependencies on $-\omega^2$, not $+\omega^2$.
Instead of (\ref{LIei}), we have now
\be
\begin{array}{ccc}
{\rm tensors} & \frac{(d+1)(d-2)}{2} & \lambda_{gr} = -\omega^2 + m_2^2 \\ &&\\
{\rm Stueckelberg\
vector} \!\!
& d-1 & \lambda_{vec} = m_1^2 \\ &&\\
{\rm scalars} &  2 &
\lambda_{sc}^\pm = \frac{
m_2^2 - m_0^2 - (d-1)m_3^2 + (d-2)\omega^2
\pm\sqrt{
\Big(m_2^2 + m_0^2 - (d-1)m_3^2 + (d-2)\omega^2\Big)^2 +\
4(d-1)m_4^4}}{2}
\end{array}
\label{LVeiRF} \ee Of the four crossings at $P1, P2, P3, P4$ in
Fig.\ref{nonperturbed}a no one is protected by the Lorentz invariance,
still in the rest frame only one is resolved by switching on
$m_4\neq 0$, Fig.\ref{nonperturbed}b.
The remaining crossings are resolved when we also
switch on non-vanishing $\vec k^2$, then
\be
\begin{array}{ccc}
{\rm tensors} & \frac{d(d-3)}{2} & \lambda_{gr} = -\omega^2 +
\vec k^2 + m_2^2 \\ &&\\
{\rm vectors}
& \ \ \ 2\times(d-2)\ \ \  & \lambda_{vec}^\pm  =
\frac{-\omega^2+\vec k^2+m_1^2+m_2^2 \pm
\sqrt{(-\omega^2+\vec k^2)^2+2(m_1^2-m_2^2)(\omega^2+\vec k^2) +
(m_1^2-m_2^2)^2  } }{2} \\ &&\\
{\rm scalars} &  4 &
C_4(\lambda_{sc}) = 0
\end{array}
\label{LVei} \ee where the polynomial $C_4$ of degree 4 in $\lambda$
is explicitly presented in Appendix III, eq.(\ref{C4}). The result is
shown in Fig.\ref{nonperturbed}c. In these pictures, the on-shell
conditions for propagating particles correspond to intersections
with the abscissa axis. Propagating particles disappear from the
spectrum when this axis coincides with one of the eigenvalue asymptotics
(thin lines in figures).

In order to investigate the DVZ jumps in the instantaneous-interaction
potential, one should instead look at Fig.\ref{kDepend},
where the four scalar $\lambda$'s are plotted as functions of
$\vec k^2$ at vanishing $\omega^2$.
Any crossing with the abscissa axis in this picture corresponds
to a tachyon.
The DVZ jumps occur when any of the eigenvalue asymptotics
(thin lines) coincide with the abscissa axis.
For more
pictures, describing the
emerging phases, see Appendix IV.

\paragraph{10. Conserved currents and instantaneous interaction.}
In gauge-invariant theories the currents, attached to
the gauge field in (\ref{pf}), are conserved: if not imposed
"by hands", this
condition appears automatically from integration over
the pure gauge degrees of freedom.
When the gauge invariance is explicitly broken, say, by the mass
terms in the second line of (\ref{lingra}), this requirement
is no longer enforced by the theory itself, instead it is
imposed on massive gravity on phenomenological grounds:
according to the currently dominating paradigm one is allowed
to "spoil" properties of the gravity sector, but not of
the matter one, which is believed to be under a much better
experimental control.

Conservation of currents is extremely important,
because it {\it de facto} eliminates some would-be propagating
degrees of freedom from the physically relevant quantity
$\Pi(k)$. Let us remind that in ordinary photodynamics,
i.e. the Maxwell theory with Lagrangian $F_{\mu\nu}^2$,
we have $\Pi = \frac{J_\mu J^\mu}{\omega^2-\vec k^2}$
what is actually equal to
\be
\Pi = \frac{J_\bot^2}{\omega^2-\vec k^2} +
\frac{J_0^2}{\vec k^2}
\ee
for conserved current, satisfying $\omega J_0 = \vec k\vec J
= |\vec k|J_{||}$, so that the longitudinal photon is actually
eliminated from $\Pi$, being substituted by a non-propagating
instant Coulomb interaction.
This fact persists in other theories with conserved currents,
including (\ref{lingra}),
even if the gauge symmetry is violated by mass terms:
the Coulomb interaction becomes the Yukawa one or even more complicated
but continue to possess an instantaneous component.
However, this is not explicitly seen at the level of eigenvalue
analysis that we performed in this paper.
This "drawback" can probably be cured by considering the
"Euclidean eigenvalues", suggested in \cite{MMMMT}, but
this can also be considered as a rather artificial trick.

Another important remark is that even if some mode drops
away from $\Pi$ when the currents are conserved, this by no means implies
it can not be {\it radiated}
(emitted) by a conserved current: one can easily imagine
situations (construct models) when a mode is emitted, but can
not be captured by another conserved current later.
This happens if space-time transverse modes are mixed with the
pure gauge ones -- what can not be generically forbidden in
gauge-violating theories. If this happens, then the fact that
the mode drops away from $\Pi$ is not sufficient to claim that it
is indeed non-propagating, and one should be careful and not
overlook such possibility.

\paragraph{11. Conclusion.}
To conclude, we used the currently popular
example of linearized massive gravity \cite{RT}
to illustrate the general  behavior
of normal modes (quasiparticles) over moduli spaces of
sophisticated physical theories
and proposed to analyze this behavior by the
standard techniques of linear and non-linear
algebra \cite{DoM}.
Already in this relatively simple example we observe
a rich pattern of bifurcations and a need to resolve
singularities in the moduli space in order to avoid
the DVZ discontinuities \cite{DVZ} and other pathologies.
This simple exercise can serve as an elementary
introduction to the general string theory problems from
the perspective of ordinary -- and even phenomenologically
acceptable -- classical field theory.
At the same time, this analysis can help to visualize
and systematize the results of \cite{RT} about the
ghost-free versions of massive gravity and further
clarify role of the Lorentz violation in constructing
such a theory, at least, at the level of quadratic
approximation.
Whatever will be its relevance for phenomenological
application, massive gravity looks very convenient for
fighting prejudices of previous experience, unapplicable
when the gauge and Lorentz invariances are broken,
and it will play a role in building new bridges between
elementary particle physics and generic quantum field/string
theory.

\section*{Acknowledgements}

We are indebted for hospitality and support to Prof.T.Tomaras and
the Institute of Theoretical and Computational Physics of
University of Crete during the summer of 2008, where this work was done.
We are specially grateful to T.Tomaras for interest, long discussions and collaboration.
We are also indebted to T.Mironova for help with the pictures.

Our work is partly supported by Russian Federal Nuclear Energy
Agency, by the joint grants 09-02-91005-ANF and 09-01-92440-CE,
by the Russian President's Grants of
Support for the Scientific Schools NSh-3035.2008.2
(A.Mir.,Al.Mor.) and NSh-3036.2008.2 (S.Mir.,An.Mor.), by RFBR grants
07-02-00878 (A.Mir.), 08-02-00287 (S.Mir.), 07-02-00645
(Al.Mor.) and 07-01-00526 (An.Mor.).

\newpage

\section*{Appendix I. Breaking Lorentz invariance in vector theory}

Here we consider the theory of massive vector field with the Lorentz invariance manifestly broken.

\subsection*{Euclidean eigenvalues}
If one chooses the
unit (Euclidean) matrix for $I$ in (\ref{I}), one has to diagonalize the
following kinetic operator:
\be\label{K}
K_{\mu\nu}=
\left(\begin{array}{ccc}
 k_{||}^2+M_0^2 & \omega k_{||} & 0\\
\omega k_{||} & \omega^2 - M_1^2 & 0 \\
0 & 0 & \omega^2-( k_{||}^2+M_1^2)
\end{array}\right)
\ee
where the spatial momentum, $k_{||}$ is directed along the first direction and
$M_{0,1}$ are the massive terms that manifestly break the Lorentz invariance. The problem of diagonalizing
this matrix leads to Euclidean eigenvalues, the result reads:
\be
\lambda_-= \frac{1}{2}\left(\Delta-\sqrt{\Delta^2-4(M_0^2\omega^2-M_1^2k_{||}^2-M_0^2M_1^2)}\right)\\
 \lambda_+ = \frac{1}{2}\left(\Delta+\sqrt{\Delta^2-4(M_0^2\omega^2-M_1^2k_{||}^2-M_0^2M_1^2)}\right)
\ee
with
\be
\Delta=\omega^2+k_{||}^2+M_0^2-M_1^2
\ee
and all other $d-2$ eigenvalues are equal to
\be
 \lambda_i=\omega^{2}-k_{||}^{2}-M_1^2
\ee
There are two kinds of dispersion laws. The condition $\lambda_i=0$ evidently leads to
the $d-2$ excitations with the dispersion law
\be\label{1}
\omega^2=k_{||}^2+M_1^2
\ee
At the same time,
the conditions $\lambda_{\pm}=0$ have only one solution
\be\label{2}
\omega^2=M_1^2+{M_1^2\over M_0^2}k_{||}^2
\ee
The simplest way to see this is to look at the
determinant of $K_{\mu\nu}$ which is equal to
\be
\left(\omega^2-k_{||}^2-M_1^2\right)^{d-2}\left(-M_0^2\omega^2+M_0^2M_1^2+M_1^2k_{||}^2\right)
\ee

\bigskip

\noindent
Now one can easily analyze these eigenvalues for physical effects:


\paragraph{tachyons:}
Dispersion law (\ref{1}) leads to a tachyon as soon as $M_1^2<0$. At the same time,
dispersion law (\ref{2}) leads to a tachyon when $M_0^2<0$.

\paragraph{superluminal:}
This may come only from dispersion law (\ref{2}), which always violates Lorentz invariance
unless $M_0^2=M_1^2$ (since then some of the vector field modes propagate with the speed of light,
and some with the speed of light times $M_1/M_0$), and, in the case of $M_1/M_0>1$,
describes the superluminal.

\paragraph{ghosts:}
The ghost content of the theory is controlled by the derivatives ${\partial\lambda\over\partial
\omega^2}$ on mass shell (i.e. at points, where $\lambda=0$). These are
\be
{\partial\lambda_i\over\partial\omega^2}=1
\ee
and
\be
{\partial\lambda_-\over\partial\omega^2}=
{M_0^4\over k_{||}^2(M_0^2+M_1^2) +M_0^4}
\ee
This derivative is zero only when $M_0^2$. However, the ghost content of the theory can not
change at this point, since it is $M_0^4$ that enters the numerator and the ghost never
emerges. If, however, $M_0^2+M_1^2<0$, there is also a
singular point where the derivative changes the sign and, therefore, the ghost emerges.

The condition $M_0=0$ is a counterpart of the condition $m_0=0$ in the
gravity case, in this case the "live" excitation branch comes away from the spectrum.
Another special case is $M_1=0$ when there only constant (in space) mode is present in the
spectrum. This is an analog of the $m_1=0$ condition in gravity.

\paragraph{DVZ jump:}
It is described by the derivatives ${\partial\lambda\over\partial
k_{||}^2}$ at zero frequencies. The derivatives are
\be
{\partial\lambda_i\over\partial k_{||}^2}=-1
\ee
and
\be\label{DVZ}
{\partial\lambda_\pm\over\partial k_{||}^2}=\left(\begin{array}{c}0\\1\end{array}\right)
\ee
Therefore, one of the derivatives is zero and, hence, there can be a DVZ jump.

\subsection*{DVZ jump}

We obtained that the necessary condition of the DVZ jump requiring that (\ref{DVZ}) to be zero
is fulfilled. However, it is identical zero for all values of parameters which makes the
general argument about the DVZ condition meaningless\footnote{It sounds as follows.
Suppose one considers the static potential at small values of mass parameters. Then, if the
coefficient in front of $k_{||}^2$ in the denominator of the propagator (=an eigenvalue)
is not going to zero with mass, nothing drastical happens. If, however, it goes to zero,
one needs some further inspection of the situation.
}.
Therefore, to establish if the DVZ jump is realized, one needs a closer
inspection of the interaction. The interaction with external currents is given by the term
$JK^{-1}J$ in the action, where $J$ is a column $(J_0,J_1,J_\perp)$ and the propagator is the
inverse of $K$ (\ref{K}). If one additionally requires for the currents to be conserved,
${\partial J^\mu\over\partial x^\mu}$, the interaction reads
\be\label{currents}
{J_0^2\over k_{||}^2}{k_{||}^2M_1^2-\omega^2M_0^2\over k_{||}^2M_1^2-\omega^2M_0^2+M_0^2M_1^2}
+{J_\perp^2\over \omega^2-k_{||}^2-M_1^2}
\ee
Bringing masses to zero in this expression in any order, as well as putting them first equal
(the Lorentz-invariant case) and then bringing to zero leads to the same result
reproducing the standard QED
\be
{J_0^2\over k_{||}^2}
+{J_\perp^2\over \omega^2-k_{||}^2}
\ee
If one consider a static potential in (\ref{currents}), i.e. the interaction generated by
a static external current, $J_\perp=0$ with $\omega=0$, one obtains
\be
{J_0^2\over k_{||}^2}{k_{||}^2\over k_{||}^2+M_0^2}
\ee
which also does not shows up any jumps. Therefore, there is no the DVZ jump in this case.

This is mostly due to a specific form of the interaction. Would be there a term, e.g.,
$M_0^4$ instead of $M_0^2M_1^2$ in the denominator of (\ref{currents}), there is
the DVZ jump. Moreover, would one consider not the static potential, but instead
the case when $\omega=k_{||}$ (with $J_\perp$ still zero), the limits of (\ref{currents})
would be different for different ways of bringing masses to zero:
\be
\left\{\begin{array}{cl}
\displaystyle{{J_0^2\over k_{||}^2}}\ \ \ & \hbox{if first}\ \ M_0\to 0\ \ \hbox{(coincides with the massless
QED case)}\\
\\
-\displaystyle{{J_0^2\over k_{||}^2}}\ \ \ & \hbox{if first}\ \ M_1\to 0\\
\\
0\ \ \ & \hbox{if first}\ \ M_0=M_1
\end{array}
\right.
\ee

\subsection*{Lorentz eigenvalues}

The other possible choice of the matrix $I$ in (\ref{I}) is the Lorentzian unit matrix, which means in the case
under consideration that one has to diagonalize the kinetic operator
\be
K^\mu_{\nu}=
\left(\begin{array}{ccc}
 -k_{||}^2-M_0^2 & -\omega k_{||} & 0\\
\omega k_{||} & \omega^2 - M_1^2 & 0 \\
0 & 0 & \omega^2-( k_{||}^2+M_1^2)
\end{array}\right)
\ee
Diagonalizing
this matrix leads to the Lorentz eigenvalues, the result reads:
\be
\lambda^L_-= \frac{1}{2}\left(\Delta_L-
\sqrt{\Delta_L^2+4(M_0^2\omega^2-M_1^2k_{||}^2-M_0^2M_1^2)}\right)\\
 \lambda^L_+ = \frac{1}{2}\left(\Delta_L+
 \sqrt{\Delta_L^2+4(M_0^2\omega^2-M_1^2k_{||}^2-M_0^2M_1^2)}\right)
\ee
with
\be
\Delta_L=\omega^2-k_{||}^2-M_0^2-M_1^2
\ee
and all other $d-2$ eigenvalues are equal to
\be
 \lambda^L_i=\omega^{2}-k_{||}^{2}-M_1^2
\ee
There are again the same two kinds of dispersion laws,
\be
\omega^2=k_{||}^2+M_1^2
\ee
and
\be
\omega^2=M_1^2+{M_1^2\over M_0^2}k_{||}^2
\ee
since the determinant
\be
\det K^\mu_\nu=\det\eta^{\mu\rho}\det K_{\rho\nu}=-\det K_{\mu\nu}
\ee

\bigskip

\noindent
Now one can again analyze the eigenvalues for physical effects:


\paragraph{tachyons:}
Since the dispersion laws are the same, the tachyon also emerge under the same conditions
as in the Euclidean case.

\paragraph{superluminal:}
Similarly, the condition for  superluminals to emerge are the same.

\paragraph{ghosts:}
The derivatives ${\partial\lambda\over\partial
\omega^2}$ on mass shell for the Lorentz eigenvalues are
\be
{\partial\lambda^L_i\over\partial\omega^2}=1
\ee
and
\be
{\partial\lambda^L_-\over\partial\omega^2}=
{M_0^4\over k_{||}^2(M_0^2-M_1^2) +M_0^4}
\ee
Again, the ghost content of the theory may change only at $M_0^2=0$ (but does not change at
this point) or when $M_0^2-M_1^2<0$.
The second condition is different for the Lorentz and Euclidean eigenvalues, while the first one
is the same. Moreover, the value of ${\partial\lambda\over\partial
\omega^2}$ is the same in both cases provided $M_1=0$ (the counterpart of $m_1=0$
condition in the gravity theory).

\paragraph{DVZ-jump:} The derivatives ${\partial\lambda\over\partial
k_{||}^2}$ on mass shell are now
\be
{\partial\lambda^L_i\over\partial k_{||}^2}=-1
\ee
and
\be
{\partial\lambda_\pm\over\partial k_{||}^2}=\left(\begin{array}{c}0\\1\end{array}\right)
\ee
Therefore, again both at $M_0=0$ and $M_1=0$ there is the DVZ jump.

Now, the lesson is that the Lorentz and Euclidean eigenvalues give the same dispersion laws and,
therefore, the same superluminal and tachyon conditions. Moreover, at least, at the case under
consideration they gives rise to the same DVZ-jump condition. The only difference is in the
ghost content conditions. However, even these latter are same provided $M_1=0$ in the vector
theory case, or $m_1=0$ in the gravity case.

Note that the Lorentz eigenvalues are often simpler to use, especially in the Lorentz-invariant
theories ($M_0=M_1$ in formulas above).
Indeed, in the latter case the eigenvalues becomes functions of only the combination
$-\omega^2+k_{||}^2$ and can be calculated at the rest frame (where $k_{||}=0$). This
simplifies calculations much, and if the ultimate results coincide, one would prefer to use
exactly the Lorentz eigenvalues.

\section*{Appendix II. Ghosts and tachyons}

In this appendix we very briefly comment on the terminology,
used in s.5 of the main text.

\bigskip

{\bf Ghosts.}
A typical example of ghost emerges in the theory of
a vector field with the Lagrangian
\be
-(\partial_\mu A_\nu)^2 = -\dot A_0^2 + \dot{\vec A}^2
- \vec k^2 (A_0^2-\vec A^2)
\label{vecgho}
\ee
Clearly, $A_0$ has a "wrong" sign in front of the kinetic
term and thus energy is unbounded from below.
This means that there can be problems in constructing
a full set of normalized states, or -- if the theory is
adequately modified at the strong-field regime --
with making the answers independent of this kind of
modification.
Such problems are typical for ghosts and one can naturally
wish to see when they can arise.
At the same time, there is nothing bad seen in the
\textit{spectrum} of the theory, if we define it with the
help of the Lorenz-invariant metric $I_L$:
$(\partial^2_{\mu\nu} - \lambda \eta_{\mu\nu})A^\nu = 0$
provides the same spectrum $\lambda = -\omega^2+\vec k^2$
for all components $A^\nu$.
If one wants to trace this type of ghosts already at the
level of spectral study, one should better use Euclidean
eigenvalues with $\eta_{\mu\nu}$ substituted by
$\delta_{\mu\nu}$, as suggested in \cite{MMMMT}.

However, example (\ref{vecgho}), though standard, is not
fully representative. Ghost appears here due to the
vector nature of the field $A_\mu$, but this does not mean
that \textit{all} ghosts should emerge for \textit{this}
reason only.
However, if the \textit{raison-d'etre} is different,
switching from Lorentzian to Euclidean eigenvalues does not
help, as we also saw in \cite{MMMMT} in analysis of the
scalar ghosts. Actually, what matters in a complicated theory
is not a particular criterium used to identify ghosts,
what matters is how normal particles turn into ghosts
and/or back to normal as one moves around in the moduli
space ${\cal M}$, and to see this many different criteria
can be used. In s.5 we mentioned the simple criterium
(\ref{ghoco}), where in the case of scalars one can use both
Lorentzian and Euclidean eigenvalues, as we do in the
present paper and in \cite{MMMMT} respectively, so that
one can compare the results.
As well as we can judge, criterium (\ref{ghoco}) is also in accord
with \cite{RT}.

\bigskip

Though this has no direct relation to content of the
present paper, in this Appendix we also remind briefly what
is bad (or good) about ghosts and tachyons.
Usually one is afraid of ghosts for the three
reasons:

(i) they can grow in time,

(ii) they can have negative norms
and thus violate perturbative unitarity,

(iii) they interact badly with normal particles.

Actually, these reasons seem different and not obligatory related to
each other.

\bigskip

The first reason (i) is pure classical:
if one begins with the Lagrangian $\alpha\dot\phi^2-V(\phi)$ with a
positive potential $\phi$ and then change the sign of $\alpha$
from positive to negative, then one immediately
and in accordance with criterium (\ref{ghoco}) obtains
solutions with imaginary frequencies, which either grow
or decrease in time, instead of oscillating.

It deserves noting that within the standard framework of QFT
perturbation theory one deals only with \textit{decreasing} or oscillating solutions.
Indeed, if one has a free
relativistic particle with the action \be \int \Big(-\omega^2 +
\varepsilon^2(\vec k)\Big) |\phi(\vec k)|^2 d\omega d\vec k \ee one
got used to define the Feynman ("casual") propagator as follows: \be
\int \frac{e^{i\omega t}d\omega} {\omega^2 - \varepsilon^2(\vec k) -
i0} = \frac{1}{2\varepsilon(\vec k)}\int \left(\frac{1}{\omega -
\varepsilon(\vec k) - i0} - \frac{1}{\omega+\varepsilon(\vec k) +
i0}\right) e^{i\omega t}d\omega =
\frac{\theta(t)e^{i\varepsilon(\vec k)t} +
\theta(-t)e^{-i\varepsilon(\vec k)t}}{2\varepsilon(\vec k)}
\ee
which is interpreted as a particle with the dispersion rule
$\omega = +\varepsilon(\vec k)>0$ propagating forward in time
and antiparticle with $\omega = -\varepsilon(\vec k)<0$
propagating backwards in time.
Technically, the integration contour is
closed by adding an infinitely remote semicircle in the upper
half-plane (with ${\rm Im}\ \omega >0$) for $t>0$ and in the lower
half-plane  (with ${\rm Im}\ \omega <0$) for $t<0$.
Accordingly contributing are different items, with poles
lying in the upper and lower half-planes respectively.
This very fact implies that the propagator \textit{can not}
grow with time at $t>0$ and \textit{can not} grow backwards in time
at $t<0$: exponents are never positive.
One should also take into account different orientations of
closed contour in two cases, and factor $2\pi i$ is included
into the definition of integral.

\begin{figure}\begin{center}
{\includegraphics*[width=0.5\textwidth]{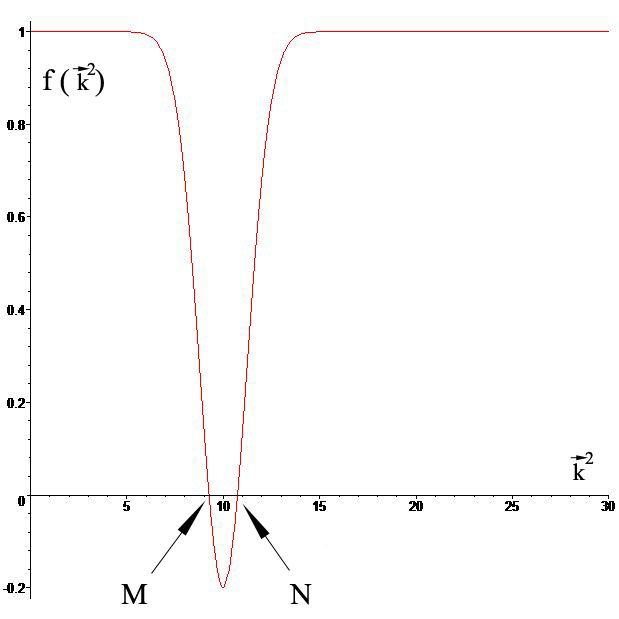}}
\caption{\footnotesize{A typical example of function $f(\vec
k^2)$ in Lorentz-violating dispersion relation $\ \lambda = -f(\vec
k^2)\omega^2 + \vec k^2 +m^2=0.\ $ For most values of space momenta
we have just an ordinary relativistic particle, while in some region
in $\vec k$ space it becomes superluminal, then carries an {\it
instantaneous} interaction (just like Newton-Coulomb-Yukawa
potential) -- this happens at points $M$ and $N$ -- and between
$M$ and $N$ it behaves like ghost, i.e. describes advanced rather
than retarded interaction.}} \label{ghost1}
\end{center}\end{figure}

If we now consider a more general action
\be
\int \Big(-f(\vec k)\omega^2 + g(\vec k)\Big)
|\phi(\vec k)|^2 d\omega d\vec k
\ee
where $f(\vec k)$ and $g(\vec k)$ can become negative
at some values of $\vec k$ (see Fig.\ref{ghost1}),
then the same propagator becomes more involved,
but exponential growth does never occur:
\be
\int \frac{e^{i\omega t}d\omega}
{f(\vec k)\omega^2 - g(\vec k) - i0} =
\left\{\begin{array}{ccc}
\frac{\theta(t)e^{it\sqrt{g/f(\vec k)}} +
\theta(-t)e^{-it\sqrt{g/f(\vec k)}}}{2\sqrt{fg(\vec k)}}
&\ {\rm when}
& \begin{array}{c} f(\vec k)>0 \\ g(\vec k)>0 \end{array}\\
&&\\
-\frac{\theta(t)e^{-t\sqrt{\left|g/f(\vec k)\right|}} +
\theta(-t)e^{t\sqrt{\left|g/f(\vec k)\right|}}}
{2\sqrt{\left|fg(\vec k)\right|}}
&\ {\rm when}
& \begin{array}{c} f(\vec k)<0 \\ g(\vec k)>0 \end{array}\\
&&\\
\frac{\theta(t)e^{-t\sqrt{\left|g/f(\vec k)\right|}} +
\theta(t)e^{-t\sqrt{\left|g/f(\vec k)\right|}}}
{2\sqrt{\left|fg(\vec k)\right|}}
&\ {\rm when}
& \begin{array}{c} f(\vec k)>0 \\ g(\vec k)<0 \end{array}\\
&&\\
\frac{\theta(t)e^{-it\sqrt{g/f(\vec k)}} +
\theta(-t)e^{it\sqrt{g/f(\vec k)}}}{2\sqrt{fg(\vec k)}}
&\ {\rm when}
& \begin{array}{c} f(\vec k)<0 \\ g(\vec k)<0 \end{array}
\end{array}\right.
\ee
Thus, we see that the choice of retarded and advanced
Green functions is defined by the {\it relative signs} in front
of different components of kinetic terms {\it with respect to}
auxiliary term $i\epsilon ||\phi(\vec k)||^2$, added to the action.
Hence, this choice depends on the definition of the norm
for the field.
In this paper we assume that these norms are defined by
Minkowski metric $\eta_{\mu\nu}$, namely for the gravity field
$||h||^2 = h_{\mu\nu}h_{\alpha\beta}
\eta^{\mu\alpha}\eta^{\nu\beta} = h_{\mu\nu}h^{\mu\nu}$.
In \cite{MMMMT} we used instead a Euclidean norm
$||h||^2_E = \sum_{\mu,\nu=0}^{d-1} h_{\mu\nu}^2$.
The disadvantage of Lorentzian norm is that by using it we allow
existence of fields with negative norms "by hands", what does
not happen with the Euclidean choice, which therefore can be
better for analysis of the physical properties of the theory
{\it per se} -- as we claimed in \cite{MMMMT} -- already for
this simple reason.
However, in Lorentz-invariant theories (like that of gauge-violating
massive vectors) the Lorentzian choice looks more "natural" (preserving Lorentz
invariance), while Lorentz-violation is introduced as a
{\it deformation} of Lorentz-invariant model and thus
does not allow an abrupt switch from Lorentzian to Euclidean
norm. Since in this paper we are more concentrated on
formal approaches to the study of eigenvalue "bundle"
over the moduli space of theories than on the physical
properties of massive gravity, we perform all analysis in terms
of Lorentzian norms and eigenvalues.
For analysis of Euclidean
norms and eigenvalues -- which we think is more physically
justified -- see \cite{MMMMT}.

\bigskip

Another typical example of the problem (ii) is well familiar
from conformal field theory: Virasoro descendants of some
"good" states can easily happen to possess negative norms
(and one tries to get rid of them in construction of
"unitary" CFT models).
In fact here, like in the previous example of negative
norms for the zero-component of a vector field,
the problem arises only when one imposes a symmetry
requirement on the metric in the space of fields.
In CFT example one requires that the Virasoro
operators are Hermitian w.r.t. the scalar product,
induced by the norm.
However, when symmetries are explicitly violated,
already at the classical level -- that of the Lagrangian --
there is no need to impose any type of symmetry requirements on
the norms and measures (which define quantization rules),
one can always use positively
definite norms on entire field
space, and in the limit where symmetries are restored
and ghosts decouple the two procedures -- with invariant
and non-invariant norms -- are actually equivalent,
though sometime it can be somewhat tedious to demonstrate.
Thus it is not so simple to decide what really happens,
either unitarity is broken by existence of negative-norm states
or, more probably, an \textit{anomaly} occurs:
the symmetry and excitation spectrum of
emerging theory is different from the one that one could
naively expect.

\bigskip

The really serious problem can be the third one, (iii). The simplest
example is again that of $A_0$-component of the vector field. In
order to have all norms positive in this theory one needs to define
vacuum state as annihilated by \textit{annihilation} operators
$\widehat{\vec a} |0> = 0$ for all the spatial components $\vec A$
of the vector field, and by \textit{creation} operator $\hat a_0 |0>
= 0$ for $A_0$. This means that one actually looses a possibility to
define a Lorentz-invariant vacuum. This could still be tolerated,
but the situation becomes dramatically worse when one tries to
switch on interactions. To the best of our knowledge so far no
consistent \textit{perturbation} theory was developed for
interacting ghosts and normal particles (though one can easily
believe that there are problem-free non-perturbative theories of
this kind: say when globally energy is positively defined while one
begins with a perturbation theory around a local maximum).

\bigskip

{\bf Tachyons.} This term is unfortunately used in total
contradiction with its literal meaning. Lexically,
"tachyon" referred to superluminal propagation, but
today we have to use the word "superluminal" for such
particles, because "tachyon is used to mean something else.
Actually, tachyon occurs when vacuum is perturbatively
unstable, then it can start to decay independently
at casually-disconnected points and this can \textit{look like}
a propagation of a superluminal excitation, but physically
the reason is obvious and very different from real
superluminals.
From the spectral point of view, what is now called tachyon
is the pole in the propagator, occurring at vanishing
frequency $\omega = 0$. The archetypical example is
the case $f(\vec k)=1$, $g(\vec k) = \vec k^2-m^2$
with the "wrong" sign in front of the mass term.
Then, at small $t$
\be
\int \frac{e^{-i\vec k\vec x}d^3 kd\omega}
{\omega^2 - k^2+m^2 - i0}\sim {m^2\over r}\left[J_1(mr)+N_0'(mr)\right]
\ee
where $J_k$ and $N_k$ are the Bessel and Neumann cylindric functions correspondingly, and the prime
means the derivative w.r.t. the argument. Therefore, in this case the propagator behaves as
$e^{imr}/r$ at large distances (small $\vec k$, there is no pole), and is singular,
$1/r$ at small distances (large $\vec k$, there is a pole). The indication of a tachyon is
non-decaying correlation at infinity. Note that simultaneously at large distances the time
correlation exponentially decays and at small distance does not. This means that at large
distances the time and spatial coordinates interchanged, and the causality is violated
(correlations do not fall outside the light cone).

\section*{Appendix III. Analysis of a model characteristic equation}

In order to illustrate the eigenvalue behavior it is instructive to
examine a model characteristic equation which is quadratic but not quartic in
$\lambda$:
\be C(\lambda) = \lambda^2 +
(2k^2+\alpha)\lambda + (\beta k^2+\gamma) = 0 \label{Cexa} \ee
In this case, one can explicitly solve all equations, and we use this to illustrate
the way the information can be extracted from plots (which are equally available beyond the
quadratic case). The
two eigenvalues are \be \lambda_\pm = \frac{-(2k^2+\alpha) \pm
\sqrt{D}}{2},\ \ \ \ \ D = (2k^2+\alpha)^2-4(\beta k^2+\gamma) =
(2k^2+\alpha-\beta)^2 +(2\alpha\beta-\beta^2-4\gamma) \label{abgei}
\ee Their $k^2$-derivatives are equal to \be \frac{\partial
\lambda_\pm}{\partial k^2} = -1 \pm
\frac{2k^2+\alpha-\beta}{\sqrt{D}}, \ee so that \be \frac{\partial
\lambda_+}{\partial k^2}\cdot \frac{\partial \lambda_-}{\partial
k^2} = \frac{D-(2k^2+\alpha-\beta)^2}{D} =
\frac{(2\alpha\beta-\beta^2-4\gamma)}{D} \ee while the resultant in the numerator of
(\ref{resra}) is
\be {\rm
resultant}_\lambda\left(C(\lambda), \frac{\partial
C(\lambda)}{\partial k^2}\right) = \det\left(\begin{array}{ccc}
1 & 2k^2+\alpha & \beta k^2+\gamma \\
2 & \beta & 0 \\ 0 & 2  &\beta
\end{array}\right) = -(2\alpha\beta-\beta^2-4\gamma)
\ee
Denominator of (\ref{resra}) is simply $D$.

\begin{figure}
\vspace{3cm}
\begin{center}
{\includegraphics[bb= 0 0 10cm 10cm,scale=0.25]
{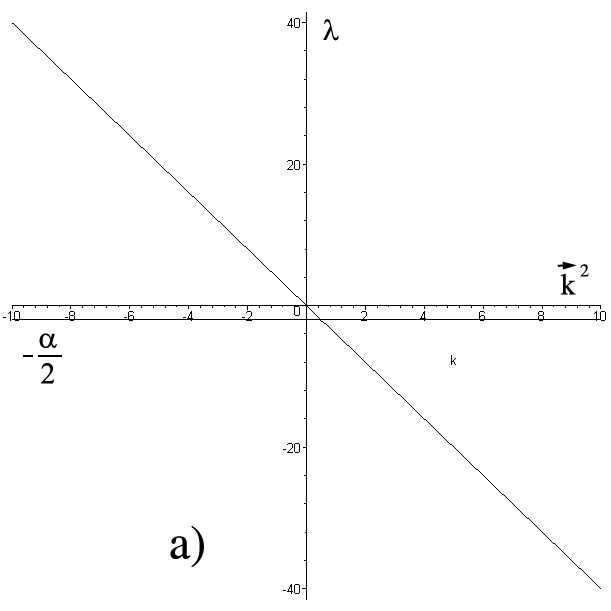}}\hspace{3cm}
{\includegraphics[bb= 0 0 10cm 10cm,scale=0.25]
{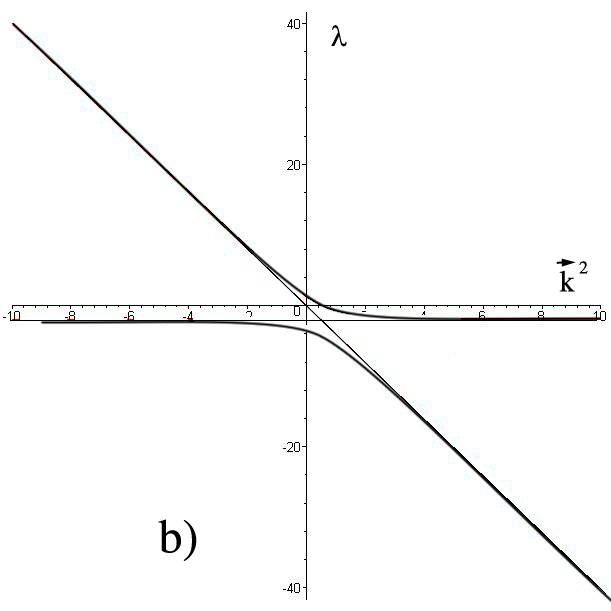}}\vspace{3cm}
\\{\includegraphics[bb= 0 0 10cm 10cm,scale=0.25]
{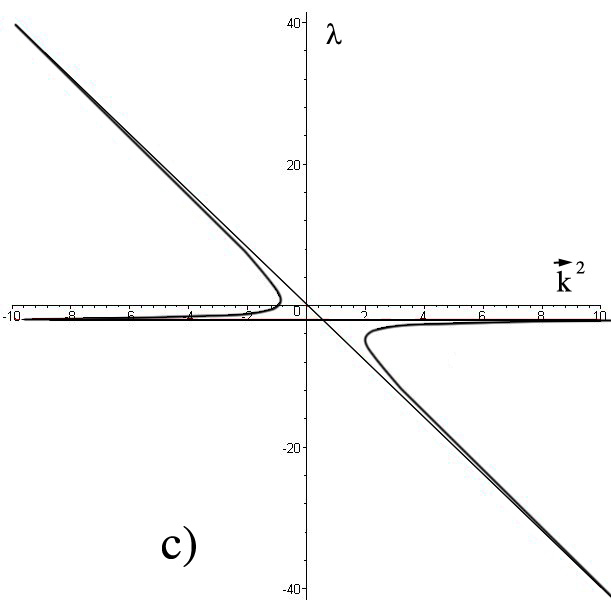}} \hspace{3cm}
{\includegraphics[bb= 0 0 10cm 10cm,scale=0.25]
{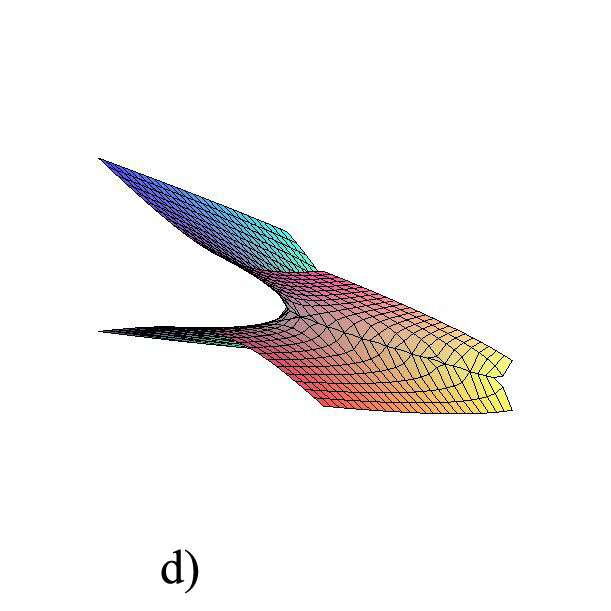}}
\caption{\footnotesize Eigenvalues (\ref{abgei}) as
functions of $k^2$ at different values of parameters $\alpha$,
$\beta$ and $\gamma$. {\bf a).} The case when
$2\alpha\beta-\beta^2-4\gamma = 0$, discriminant $D$ is a full
square and $\lambda_\pm$ become linear functions of $k^2$, or
$\lambda_\pm = \frac{-(2k^2+\alpha) \pm |2k^2+\alpha-\beta|}{2}$, to
be exact. The horizontal line is at $\lambda_\infty =
-\frac{1}{2}\beta$. \ \ \ {\bf b).} Resolution of the crossing
singularity at $2\alpha\beta-\beta^2-4\gamma > 0$, when
$\lambda_\pm$ are real at all values of $k^2$. Asymptotically at
$k^2\rightarrow \pm\infty$ eigenvalues tend to $\lambda_\infty = -
\lim_{k^2\rightarrow\infty} \frac{\beta k^2 + \gamma}{2k^2+\alpha} =
- \frac{1}{2}\beta$ (or to infinity). Punctured lines show the same
cross as in Fig.a, the role of the resolution (deformation)
parameter is played by $\gamma$. \ \ \ {\bf c).} Resolution of the
crossing singularity at $2\alpha\beta-\beta^2-4\gamma > 0$, when
$\lambda_\pm$ fail to be real-valued at some $k^2$.\ \ \ {\bf d).}
The $3d$ plot of $\lambda_\pm$ as function of $k^2$ and $\gamma$.
The saddle structure is clearly seen. It degenerates to
Fig.\ref{LIscals}a in the special case of (\ref{LIC}) when
$2\alpha\beta-\beta^2-4\gamma = 4(d-1)B^2$ and is never negative.} \label{abgplot}
\end{center}\end{figure}

This implies that $\frac{\partial \lambda}{\partial k^2}=0$ in two
cases: either when $k^2=\pm \infty$ and discriminant $|D|=\infty$ or
when $2\alpha\beta-\beta^2-4\gamma = 0$ and $D$ is a full square, so
that $\lambda_\pm$ become linear functions of $k^2$. This property
is nicely illustrated by the plots in Fig.\ref{abgplot}. In this way
one can extract information from eq.(\ref{resra}) and from plots,
what is especially useful in the realistic case (\ref{LVei}), when
$C_4(\lambda)$ has degree $4$ and analytical approach is less
straightforward.

Explicitly in the linearized Lorentz-violating gravity (\ref{lingra})
the characteristic polynomial for the scalar modes is equal to \be\label{C4}
C_4(\lambda) = \lambda^4 \ +\ \Big((d-3)(-\omega^2+\vec
k^2)+m_0^2-m_1^2-2m_2^2+(d-1)m_3^2\Big)\lambda^3 + \ee
\vspace{-0.1cm} {\footnotesize
$$
+\Big(\!\!-(d-2)(-\omega^2 + \vec k^2)^2 -\big((d-3)(m_0^2- m_1^2 -
m_2^2) - (d-1)m_3^2 \big)\omega^2 +\big( (d-3)(m_0^2 -m_1^2-m_2^2
+m_3^2)-2(d-2)m_4^2\big)\vec k^2 - $$ $$ -m_0^2m_1^2-2m_0^2 m_2^2 +2
m_1^2 m_2^2 +m_2^4  + (d-1)(m_0^2m_3^2 -m_1^2 m_3^2-m_2^2 m_3^2
-m_4^4)\Big) \lambda^2 -
$$ \vspace{0.2cm}
$$
-\Big((d-2)\big[(m_0^2-m_1^2)\omega^4 -2(m_0^2 -m_2^2
+m_3^2-m_4^2)\omega^2\vec k^2 -(m_1^2+m_2^2-m_3^2)\vec k^4\big] - $$
$$ -\big((d-3)(m_0^2m_1^2+m_0^2m_2^2-m_1^2m_2^2)
+(d-1)(m_0^2m_3^2-m_1^2m_3^2 -m_4^4)\big)\omega^2 + $$ $$
+\big((d-3)(m_0^2m_1^2+m_0^2m_2^2-m_0^2m_3^2-m_1^2m_2^2+
m_1^2m_3^2+m_4^4) -2(d-2)(m_1^2+m_2^2)m_4^2 \big)\vec k^2 - $$ $$ -
2m_0^2m_1^2m_2^2 - m_0^2m_2^4 + m_1^2m_2^4 +(d-1)(m_1^2+m_2^2)(
m_0^2m_3^2 -m_4^4) - (d-1)m_1^2 m_2^2m_3^2 \Big)\lambda +
$$ \vspace{0.2cm}
$$
+\Big( (d-2)\big[ m_0^2m_1^2\omega^4 - 2(m_0^2m_2^2 -m_0^2
m_3^2-m_1^2 m_4^2+ m_4^4  ) \omega^2\vec k^2 -
m_1^2(m_2^2-m_3^2)\vec k^4\big] - $$ $$ -m_1^2 \big((d-3)m_0^2 m_2^2
+ (d-1)(m_0^2m_3^2 -m_4^4)\big)\omega^2
+ m_1^2 \big((d-3)(m_0^2 m_2^2 -m_0^2 m_3^2 +m_4^4)  -2(d-2) m_2^2
m_4^2 \big)\vec k^2 -$$ $$ - m_0^2m_1^2m_2^4 +
(d-1)m_1^2m_2^2(m_0^2m_3^2-m_4^4)\Big)
$$}
At $\vec k = 0$, i.e. in the rest frame, the characteristic
polynomial (\ref{C4}) factorizes: \be \left.
C_4(\lambda)\right|_{\vec k =0} = (\lambda +
\omega^2-m_2^2)(\lambda-m_1^2)\cdot \nn \\ \cdot
\left\{\lambda^2+\lambda\Big(m_0^2-m_2^2+(d-1)m_3^2-(d-2)\omega^2\Big)
-m_0^2\Big(m_2^2+(d-2)\omega^2\Big)+(d-1)(m_0^2m_3^2-m_4^4)\right\}
\label{C4RF} \ee For (\ref{LIca}) the last bracket turns into \be
C_2(\lambda) = \lambda^2+\lambda\Big(dB-2A-(d-2)\omega^2\Big)
+\Big((d-2)(A-B)\omega^2+A(A-dB)\Big) \label{LIC} \ee with the two roots
given by the last row in (\ref{LIei}), \be \lambda_\pm=A -\frac{dB -
(d-2)\omega^2\pm \sqrt{ (d-2)^2(B-\omega^2)^2 + 4(d-1)B^2}}{2} \ee
Example (\ref{Cexa}) can be now used upon identification $k^2 =
-\frac{d-2}{2}\omega^2$, $\alpha = dB-2A$, $\beta = 2(B-A)$, $\gamma
= A(A-dB)$ and $2\alpha\beta-\beta^2-4\gamma= 4(d-1)B^2\geq 0$. In
general in the rest frame one gets from the last bracket in
(\ref{C4RF}): $\alpha = m_0^2-m_2^2+(d-1)m_3^2$, $\beta = 2m_0^2$,
$\gamma = (d-1)(m_0^2m_3^2-m_4^4) - m_0^2m_2^2$ and
$2\alpha\beta-\beta^2-4\gamma = 4(d-1)m_4^4 \geq 0$. This implies that the
singularity is resolved at $\omega^2=0$ in the single possible way, which
explains the universal structure of Fig.\ref{nonperturbed}b.

If instead of $\vec k =0$ we put $\omega=0$, characteristic
polynomial $C_4(\lambda)$ factorizes in a less radical way: \be
\left. C_4(\lambda)\right|_{\omega =0} = (\lambda-m_1^2)
\Big\{\lambda^3 + \Big( (d-3)\vec k^2 +m_0^2 -2m_2^2
+(d-1)m_3^2\Big)\lambda^2 + \label{Com0} \ee \vspace{-0.3cm}
$$
+\Big( -(d-2)\vec k^4 +
\big[(d-3)(m_0^2-m_2^2+m_3^2)-2(d-2)m_4^2\big]\vec k^2 + \big[m_2^4
- 2m_0^2m_2^2 +(d-1)(m_0^2m_3^2-m_2^2m_3^2-m_4^4)\big] \Big)\lambda
+
$$
$$
+\Big((d-2)(m_2^2-m_3^2)\vec k^4 + \big[(d-3)m_0^2(m_3^2-m_2^2)
+2(d-2)m_2^2m_4^2 -(d-3)m_4^4\big]\vec k^2 +
\big[m_0^2m_2^4-(d-1)m_0^2m_2^2m_3^2  + (d-1)m_2^2m_4^4\big]
\Big)\Big\}
$$
The roots of this equation are plotted in Fig.\ref{kDepend}. Of course,
in the Lorentz-invariant case (\ref{LIca}) this (\ref{Com0}) further
reduces to $(\lambda-\vec k^2-A)$ times (\ref{LIC}), with
$-\omega^2$ substituted by $\vec k^2$.

For the full $C_4(\lambda)$ in (\ref{C4}) the resultants of
$C_4(\lambda)$ with $\frac{\partial C(\lambda)}{\partial \omega^2}$
and $\frac{\partial C(\lambda)}{\partial \vec k^2}$ in the numerator
of (\ref{resra}) are rather complicated and essentially different,
however, they contain two common $d$-independent factors: \be
\Big((m_0^2+m_1^2)(m_1^2-m_2^2+m_3^2) - m_4^4\Big) \label{resfac1}
\ee and \be \Big(m_4^2\omega ^4   -(m_0^2+m_2^2-m_3^2)\omega^2\vec
k^2 -m_4^2\vec k^4
-(m_0^2m_4^2 + m_2^2m_4^2 + m_3^2m_4^2)\omega^2 + \nn \\
+(m_0^2m_3^2+m_2^2m_3^2-m_3^4-2m_4^4)\vec k^2 +
(m_0^2m_3^2m_4^2+m_2^2m_3^2m_4^2 -m_4^6) \Big) \label{resfac2} \ee
The first factor (\ref{resfac1}) vanishes in the Lorentz invariant case
(\ref{LIca}), when one of the eigenvalue lines is obligatory
horizontal. Furthermore, ${\rm resultant}_\lambda\left(C(\lambda),
\frac{\partial C(\lambda)}{\partial \omega^2}\right) \sim \vec k^2$,
while ${\rm resultant}_\lambda\left(C(\lambda), \frac{\partial
C(\lambda)}{\partial \vec k^2}\right) \sim \omega^2$, so that they
vanish at $\vec k=0$ and $\omega = 0$ respectively -- in accordance
with Figs.\ref{nonperturbed}b and \ref{LIscals}b, which both contain one horizontal
eigenvalue line (associated with the spatial Stueckelberg scalar). In
addition, in the rest frame, at $\vec k = 0$, the second factor
(\ref{resfac2}) is proportional to $m_4$, what corresponds to
appearance of the second horizontal eigenvalue line in Fig.\ref{nonperturbed}a
at $m_4=0$. As to the first factor (\ref{resfac1}), when it vanishes
beyond the Lorentz-invariant case (\ref{LIca}), it still signals
that a horizontal line occurs. In fact, this is the line $\lambda =
m_1^2$: if (\ref{resfac1}) vanishes, then $C_4(\lambda)$ is
divisible by $(\lambda-m_1^2)$ for all values of $\omega$ and $\vec
k$.


\section*{Appendix IV. Eigenvalue bundle over the moduli space}

\def\thesubsection{\arabic{subsection}}

\subsection{Deformation of four crosses}

The four-cross pattern, Fig.\ref{nonperturbed}a, describes the dependence
of four scalar eigenvalues of $-\omega^2$ when $m_4^2=0$ and $\vec
k^2=0$. Then the two horizontal lines are $\lambda_{sS}=-m_0^2$ and
$\lambda_{S}=m_1^2$, while two other lines with slopes $+1$
(normal particle) and $-(d-2)$ (ghost) are given by $\lambda_{sT} =
-\omega^2+m_2^2$ and $\lambda_{stT} = (d-2)\omega^2 + m_2^2-(d-1)m_3^2$, i.e. they
intersect the ordinate axis at $m_2^2$ and $m_2^2 -
(d-1)m_3^2$ respectively. There are two propagating (on-shell) modes
with $\lambda=0$, one is normal, another one is ghost.

When $m_4^2\neq 0$ is switched on, Fig.\ref{nonperturbed}b, one of the four crossings is
resolved: the one between $\lambda_{sS}$ and $\lambda_{stT}$. The intersecting eigenvalues
repulse but intersection with the abscissa axis  corresponds to a
ghost in both cases, $m_0^2>0$ and $m_0^2<0$: depending on the
sign of $m_0^2$ the on-shell ghost comes from either lower or
upper of the two branches. The only exception is the case of
$m_0^2=0$: then this on-shell mode simply disappears at infinity
and the ghost is eliminated, at least at $\vec k^2=0$.

Switching on $\vec k^2\neq 0$ resolves all the four crossings (even
if $m_4^2=0$). This adds one more option: that ghost can be
eliminated not only when $m_0^2=0$ but also when $m_1^2=0$,
exactly in the same way as above. However, with increasing $\vec
k^2$ and $m_4^4$ the patterns deviate pretty far from the
four-crosses of Fig.\ref{nonperturbed}a, see Fig.\ref{laplot1def},\ref{diffmass11} for some
examples. This means that even if ghosts are eliminated in the
vicinity of four-cross pattern, they can re-appear at larger values
of the momentum $\vec k^2$. Also the case of large $m_4^4$ requires
more careful analysis.

As clear from Figs.\ref{laplot1def},\ref{diffmass11}, the only two possibilities to have
ghost-free models arise at either $m_0^2=0$ or at $m_1^2=0$.

\subsection{Analysis through the chain of bifurcations}

Analysis of the properties of propagating particles can be performed
in a certain order, because actually there is a hierarchy of
interesting properties.

1) Plot $\lambda(-\omega^2)$ at fixed $\vec k^2$ and masses or
$\lambda(\vec k^2)$ at fixed $\omega^2$ and masses. Of interest are
on-shell states $\lambda=0$ and the slopes
$-\left.\frac{\partial\lambda}{\partial\omega^2}\right|_{\lambda=0}$
or $\left.\frac{\partial\lambda}{\partial\vec
k^2}\right|_{\lambda=0}$ at these points. In what follows we
consider mostly the first option: $\lambda(-\omega^2)$.

2) The signs of derivatives are actually controlled by topology of
the graph $\lambda(\omega^2)$, especially by positions of the
branching points: zeroes of discriminant $\ {\rm
discrim}\Big(C(\lambda)\Big)$ where different branches merge or
intersect. These critical points $\omega_{{\rm cr}}^2$ are
themselves {\it not} on-shell, but they actually define the
properties of on-shell modes. They depend both on $\vec k^2$  and
masses. Of primary interest is their dependence on space momentum
$\vec k^2$ at fixed masses.

3) The properties of on-shell particles change qualitatively at
bifurcation points when $\omega^2_{{\rm cr}}$ merge, vanish or go to
infinity. This is controlled by zeroes of the next-level
discriminant $\ {\rm discrim}\Big(\omega^2_{cr}(\vec k^2)\Big)$.
These zeroes $\vec k^2_{{\rm cr}}({\rm masses})$ depend only on
masses and change when the masses change, i.e. when we move along
the moduli space ${\cal M}$. At some points of ${\cal M}$ there can
be regions in momentum space where on-shell particles are ghosts,
and regions where they are always normal or are ghosts for all
values of $\vec k$.

4) The boundaries between these regions are defined by the
next-order discriminants $\ {\rm discrim}\Big(\vec
k^2_{cr}(masses)\Big)$. One can again make an iterative study:
change first some of masses, most conveniently, $m_4^2$, and then
the others, considering higher and higher order discriminants at
each step.

\subsection{Examples}

We give now examples of such hierarchical analysis.

\begin{figure}
\begin{center}
{\includegraphics[width=0.65\textwidth]
{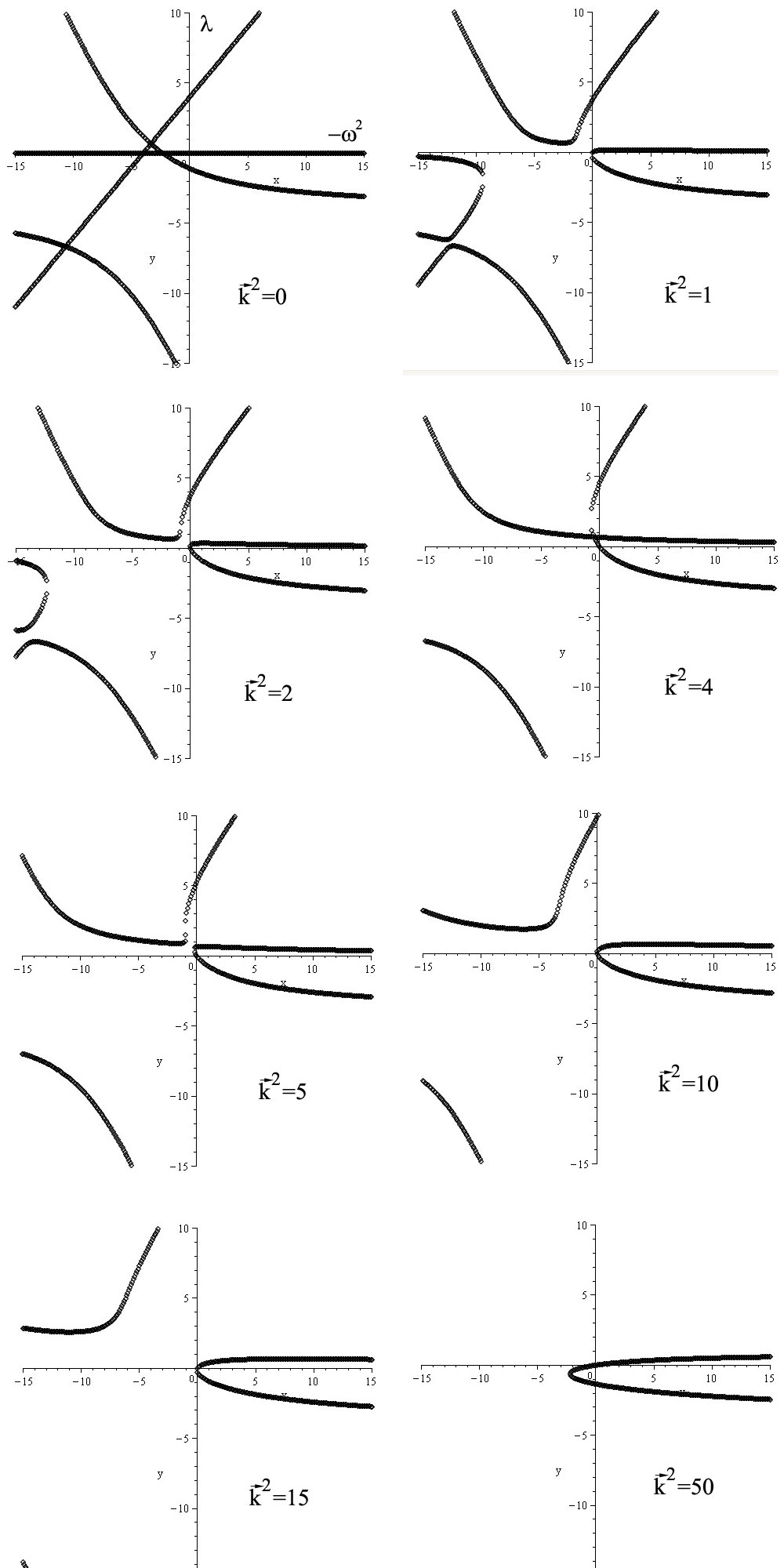}}\vspace{0.1cm} \caption{\footnotesize{ Pattern of
eigenvalues dependence on $-\omega^2$ for masses $m_0^2=4$,
$m_1^2=0$, $m_2^2=4$, $m_3^2=6$, $m_4^2=3.53553\ldots$ and different
$\vec k^2$. The slope of the curve
$\lambda(-\omega^2)$ at $\lambda=0$ (on shell) changes from positive
to negative and back to positive, thus demonstrating existence of a
window in $\vec k^2$ space where propagating particle behaves as a
ghost -- in accordance with Fig.\ref{discrC1}. }} \label{laplot1def}
\end{center}\end{figure}

\begin{figure}
\vspace{2cm}
\begin{center}
{\includegraphics[bb= 0 0 10cm 10cm,scale=0.25]
{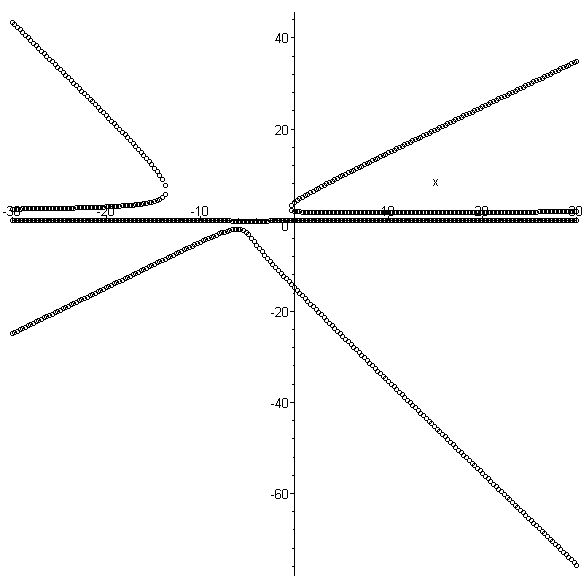}}\hspace{3cm}
{\includegraphics[bb= 0 0 10cm 10cm,scale=0.25]
{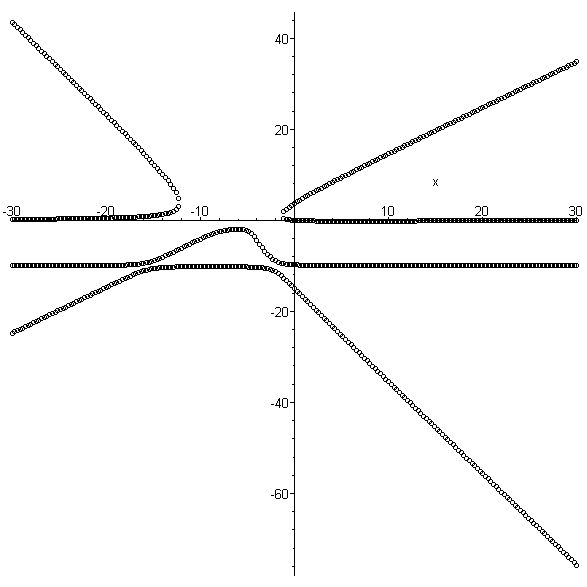}}\vspace{3cm}
\\{\includegraphics[bb= 0 0 10cm 10cm,scale=0.25]
{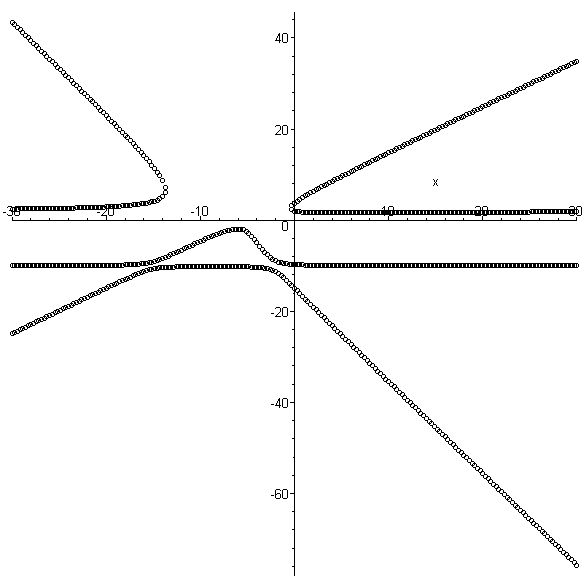}} \hspace{3cm}
{\includegraphics[bb= 0 0 10cm 10cm,scale=0.25]
{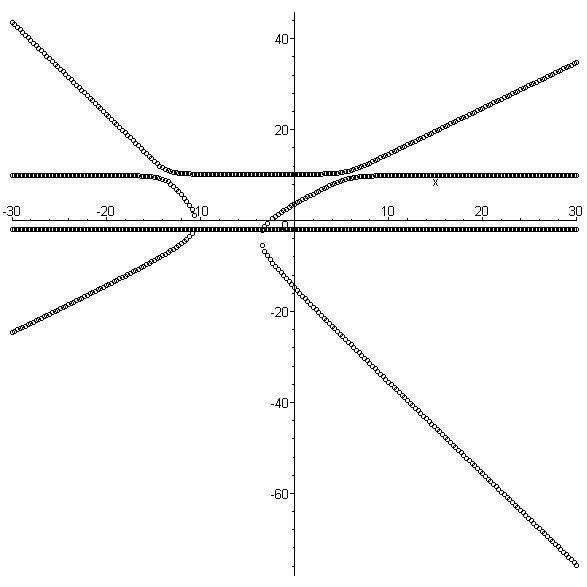}}\vspace{3cm}
\\{\includegraphics[bb= 0 0 10cm 10cm,scale=0.25]
{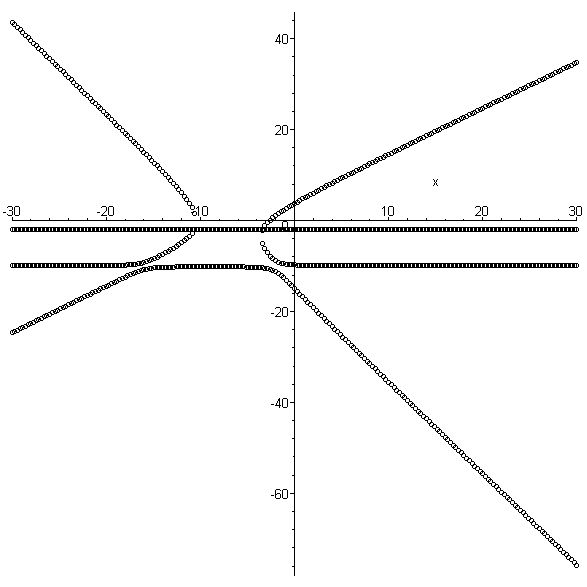}} \hspace{3cm}
{\includegraphics[bb= 0 0 10cm 10cm,scale=0.25]
{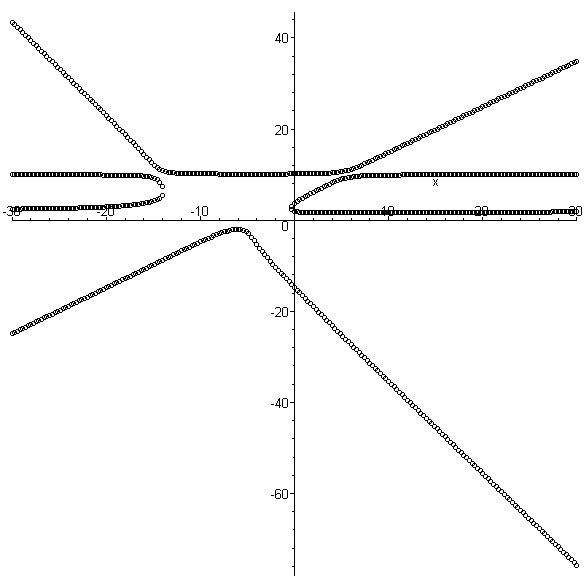}} \caption{\footnotesize{ The graph of the dependence of
eigenvalues on $\omega^2$ for $k^2=1$ and different values of masses $m_1$ and $m_0$.
Masses at the left pictures (down from above) are equal to $m_0^2=0$, $m_1^2=2$, $m_2^2=4$,
$m_3^2=6$, $m_4^2=0$; $m_0^2=10$, $m_1^2=2$, $m_2^2=4$,
$m_3^2=6$, $m_4^2=0$; $m_0^2=10$, $m_1^2=-2$, $m_2^2=4$,
$m_3^2=6$, $m_4^2=0$. Similarly, those at the right pictures are
$m_0^2=10$, $m_1^2=0$, $m_2^2=4$,
$m_3^2=6$, $m_4^2=0$; $m_0^2=-10$, $m_1^2=-2$, $m_2^2=4$,
$m_3^2=6$, $m_4^2=0$; $m_0^2=-10$, $m_1^2=2$, $m_2^2=4$,
$m_3^2=6$, $m_4^2=0$.
Therefore, the two upper pictures  correspond to the case when one of these masses is equal to
zero. }} \label{diffmass11}
\end{center}\end{figure}

\begin{figure}
\begin{center}
{\includegraphics[width=\textwidth]
{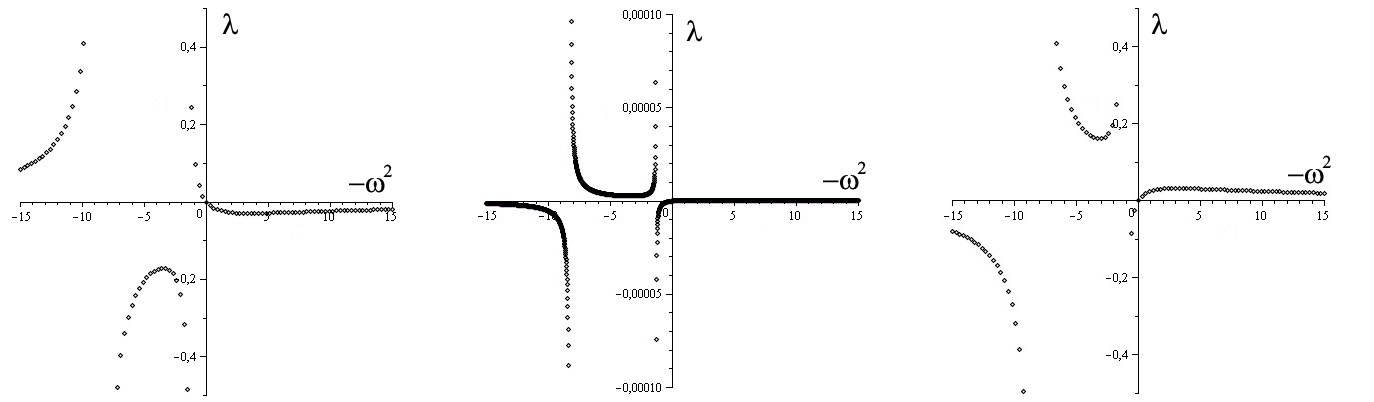}} \caption{\footnotesize{ Variation of the
eigenvalue curves in Fig.\ref{laplot1def} with the change of  mass
$m_4$: $m_4^2=3;\ 2.82843;\ 2.64575$ (from the left to the right). The scale in the middle
picture is different.}} \label{laplot2def}
\end{center}\end{figure}

1) Some plots of the four-branch function $\lambda(-\omega^2)$
are shown in Figs.\ref{laplot1def} and \ref{diffmass11}. In Fig.\ref{laplot1def}
the values of masses
are fixed and different plots are for different values of
$\vec k^2$. In Fig.\ref{diffmass11} we fix $\vec k^2=1$ instead, but
change two of the five masses ($m_0^2$ and $m_1^2$) instead.
If another mass ($m_4^2$) is changed we get a very different
pattern, Fig.\ref{laplot2def}.
There is no problem in making many more plots of this kind.
The problem is to find some reasonable way to put this
huge collection in order. This is what above hierarchical
procedure is supposed to do.

2) From Figs.\ref{laplot1def} and \ref{diffmass11} it is clear that the whole pattern
is very well controlled by position of the branching points
(where the tangent line becomes vertical).
These branching points can be defined by pure algebraic means:
they are zeroes of discriminant:
solutions to the equation
\be
{\rm discrim}_\lambda \Big(C(\lambda)\Big)=0
\ee
Discriminant at the l.h.s. is a little too long to present
here, but it is an explicit
polynomial\footnote{As a function of its coefficients discriminant
of the order-$4$ polynomial $C(\lambda)=\sum_{i=0}^4 C_i\lambda^i$
is given by
$$
-4C_4C_2^3C_1^2+16C_4C_2^4C_0-128C_4^2C_0^2C_2^2-27C_3^4C_0^2
-6C_4C_0C_3^2C_1^2+144C_4C_0^2C_2C_3^2+144C_4^2C_0C_2C_1^2
+18C_4C_3C_1^3C_2+C_2^2C_3^2C_1^2
- $$ $$ -4C_2^3C_3^2C_0-4C_3^3C_1^3
+256C_4^3C_0^3-192C_4^2C_0^2C_3C_1-80C_4C_3C_1C_2^2C_0
+18C_3^3C_1C_2C_0-27C_4^2C_1^4
$$
It is a polynomial of degree $2(4-1)=6$ in the coefficients
and can be read from the celebrated Sylvester formula or
represented as a combination of two simple diagrams,
see \cite{discres,DoM}.
Substitution of coefficients expression through masses,
frequency and momentum from (\ref{C4}) makes this expression
rather lengthy.}
and can be easily evaluated for any particular set of masses.
After that its zeroes can be found numerically
and they are plotted in the center of Fig.\ref{discrC1} as functions of $\vec k^2$
for the same values of masses that were chosen in Fig.\ref{laplot1def}.
One can easily compare Figs.\ref{laplot1def} and \ref{discrC1}, and the moral is
that essential information about the pattern in Fig.\ref{laplot1def}
is actually contained in the far simpler and pure
algebraic plot in Fig.\ref{discrC1}.
In fact, one can easily plot zeroes of the same discriminant
as functions of masses instead of $\vec k^2$ and reproduce
the essential properties of Fig.\ref{diffmass11} instead of Fig.\ref{laplot1def}.

3) Fig.\ref{discrC1} itself can be changed if we vary remaining parameters.
In Fig.\ref{discrC1} positions of the branching points were plotted
as functions of $\vec k^2$. Fig.\ref{laplot3def} shows what happens
with Fig.\ref{discrC1} when one of the mass parameters is changed.
The difference between Figs.\ref{discrC1} and Fig.\ref{laplot3def} could be
systematically controlled in an algebraic way, if we look
at zeroes of the {\it repeated discriminant} -- the crossing/merging
points of the three branches in Fig.\ref{discrC1}. This structure is discussed
below and pictured in Fig.\ref{264}-\ref{fig4}.

4) Procedure can be repeated again and again, going to higher
and higher codimension in the moduli space ${\cal M}$.
Thus we obtain a systematic approach to the study of
bifurcations/reshufflings of eigenvalue bundle and to
construction of phase diagrams of the theory.

\subsection{The bundle structure}

It is instructive to present the same in a slightly different
words -- and pictures,-- by making more explicit the structure of
\textit{eigenvalue bundle} over the moduli space of linearized
massive gravity.

Let us fix the values of four masses,
$m_0^2=4$, $m_1^2=0$, $m_2^2=4$, $m_3^2=6$
and look what happens when we change $m_4^2$
from $2.64$ to $3.53$.
The choice of masses is rather arbitrary
with two exceptions: $m_1^2$ is taken vanishing
in order to look at appearance and disappearance
of the on-shell ghost, and $m_4^2$ is chosen to
vary in the vicinity of the critical value, where
\be\label{Delta}
\Delta \equiv m_0^2(m_3^2-m_2^2) - m_4^4 = 0
\ee
and where an interesting bifurcation occurs.
In this particular case
the critical value of $m_4^2$ is
$m_0\sqrt{m_3^2-m_2^2} = 2\sqrt{2} = 2.828427\ldots$.

\begin{figure}
\begin{center}
{\includegraphics[width=\textwidth]
{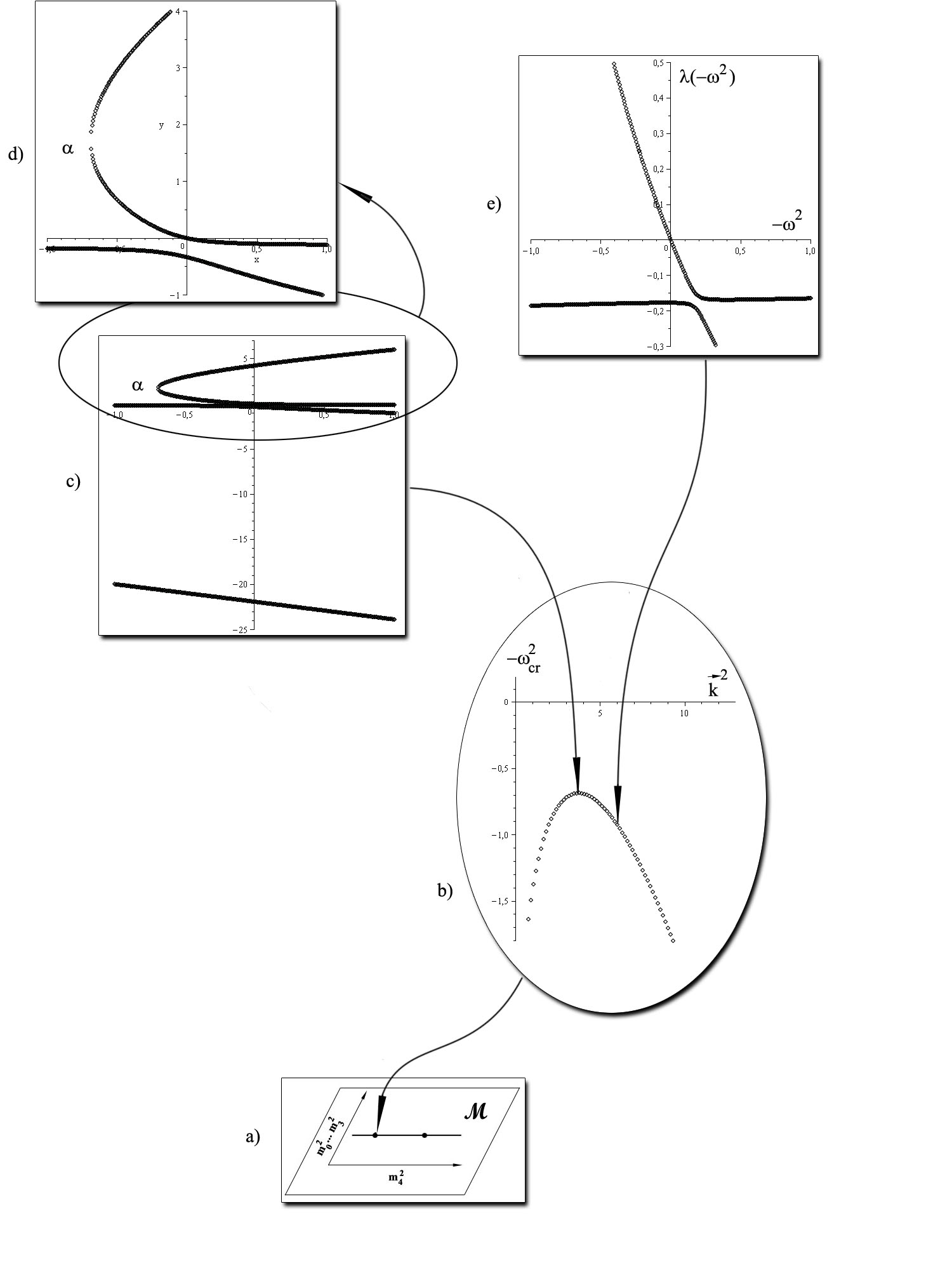}}\caption[t]{\phantom{spasmodical}}
\label{264}
\end{center}\end{figure}

Thus we begin from $m_4^2 = 2.64$. Over this point of the moduli
space ${\cal M}$ there is a fiber of our "bundle", consisting of the
4-branched function $\lambda(\omega,\vec k)$. Instead of hanging
such 3-dimensional fiber over the base point we do another thing: we
hang first a $2d$ plot of $\omega^2_{crit}(\vec k^2)$, which shows
how the three critical points $\alpha, \beta, \gamma$ -- the three
zeroes of discriminant ${\rm discrim}\Big(C(\lambda)\Big)$ -- change
with the variation of momentum $\vec k^2$. After that over each
point of \textit{this} fiber -- which we call \textit{discriminant
fiber} in what follows -- we hang the $2d$ plot $\lambda(-\omega^2)$
(the \textit{eigenvalue fiber}), as shown in Fig.\ref{264}. In this
particular case of $m_4^2 = 2.64$ discriminant fiber consists of a
single real branch and only a single branching point $\alpha$ is
seen in the  eigenvalue plot. We show also the enlarged vicinity of
$\alpha$ in accompanying figure, where one non-very-interesting
branch is not actually seen. It is clear from this picture that
there is a single on-shell scalar, it is located at $\omega^2=0$ and
it is ghost, because the slope of the branch is negative at the
intersection with abscissa axis. In fact, this is a very exotic
excitation, being simultaneously a carrier of instantaneous
interaction $\omega^2=0$ and a ghost since $d\lambda/d\omega^2<0$ on
shell. Its characteristic dispersion relation is $-\omega^2\vec
k^2=i\epsilon$ \cite{MMMMT}. Actually it remains of this same kind
for all values of $m_4^2$, only the coefficient at the l.h.s.
becomes a sophisticated function of $\vec k^2$ and can even change
sign with the variation of $\vec k^2$. The physical implications of
such "instantaneon" excitation in the spectrum remain an interesting
subject for future investigation.

Now we start increasing $m_4^2$.
At $m_4^2=2.828428\ldots$, i.e. at the critical value
$\Delta = 0$ a new couple of branches
shows up in discriminant fiber (they were complex at lower
values\footnote{
One can see that the two complex branches are already
very close to become real by looking at eigenvalue plot
at $m_4^2 = 2.64$ and $\vec k^2=6$: it is pretty clear
that something is going to happen, and this behavior of
the eigenvalue curves signals that discriminant zero is
nearby -- just not seen in the real section of generic
complex picture.
} of $m_4^2$) and they get well separated soon enough,
we show an example at $m_4^2=2.9$.
It is see that at this value of $m_4^2$ one of the two
new branches merges with the old one at $\vec k^2\approx 2$
-- and this is reflected in the properties of the
eigenvalue fiber.

\begin{figure}
\begin{center}
{\includegraphics[width=\textwidth]
{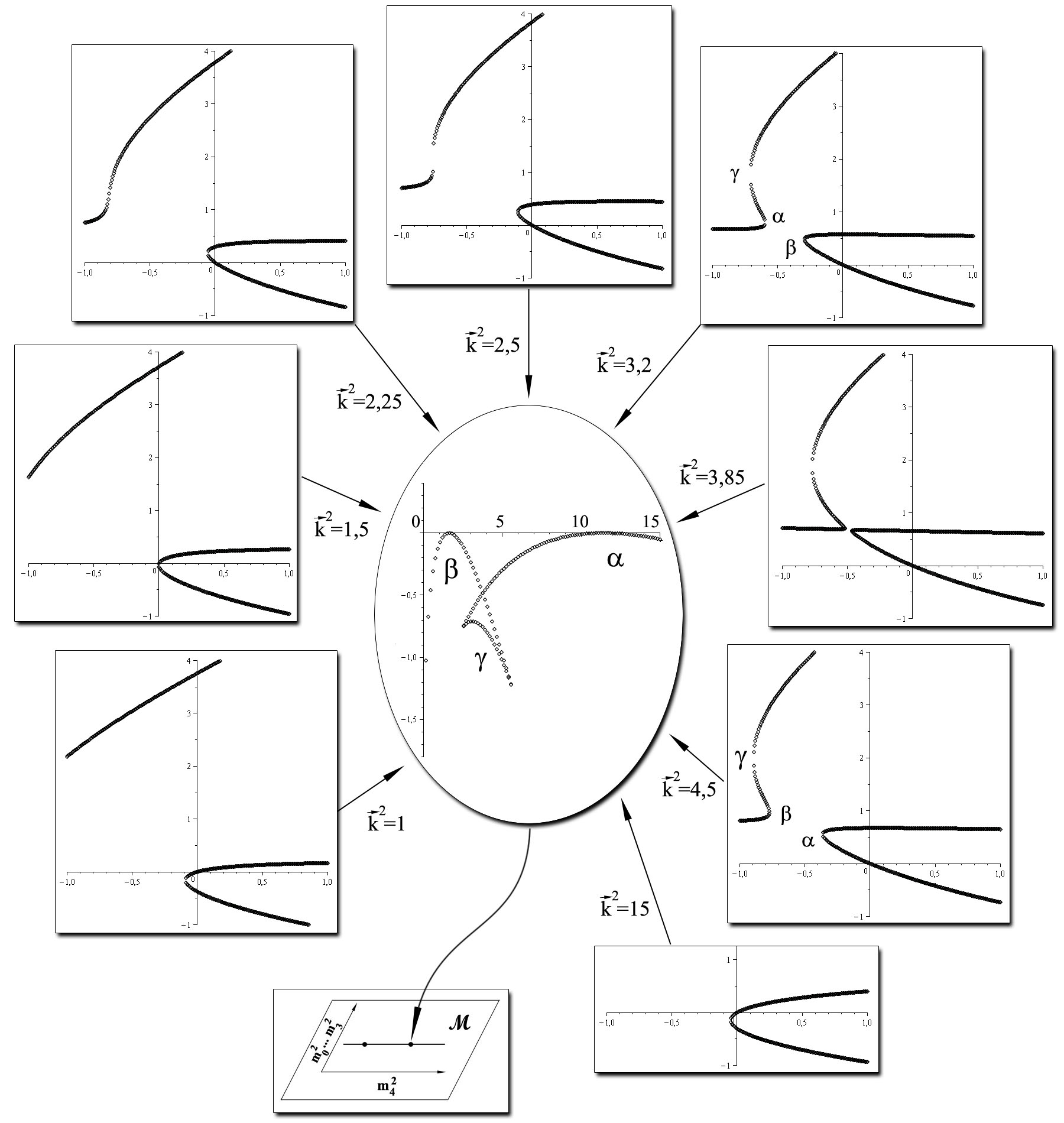}}\caption[t]{\phantom{sjdaskjdhad}}\label{discrC1}
\end{center}\end{figure}

\begin{figure}
\begin{center}
{\includegraphics[width=\textwidth]
{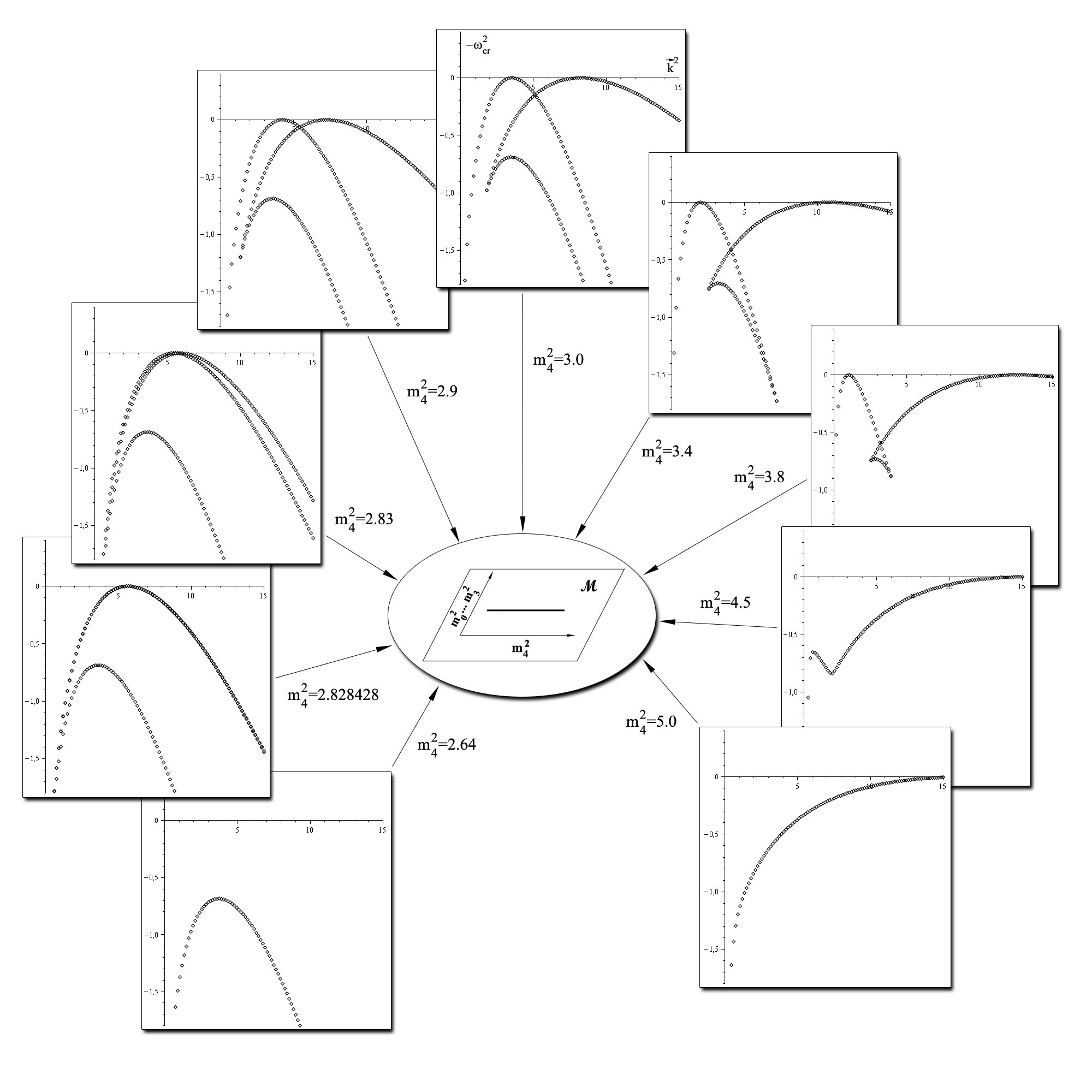}}\caption[t]{\phantom{sjdaskjdhad}}\label{laplot3def}
\end{center}\end{figure}

\begin{figure}
\begin{center}
{\includegraphics[width=\textwidth]
{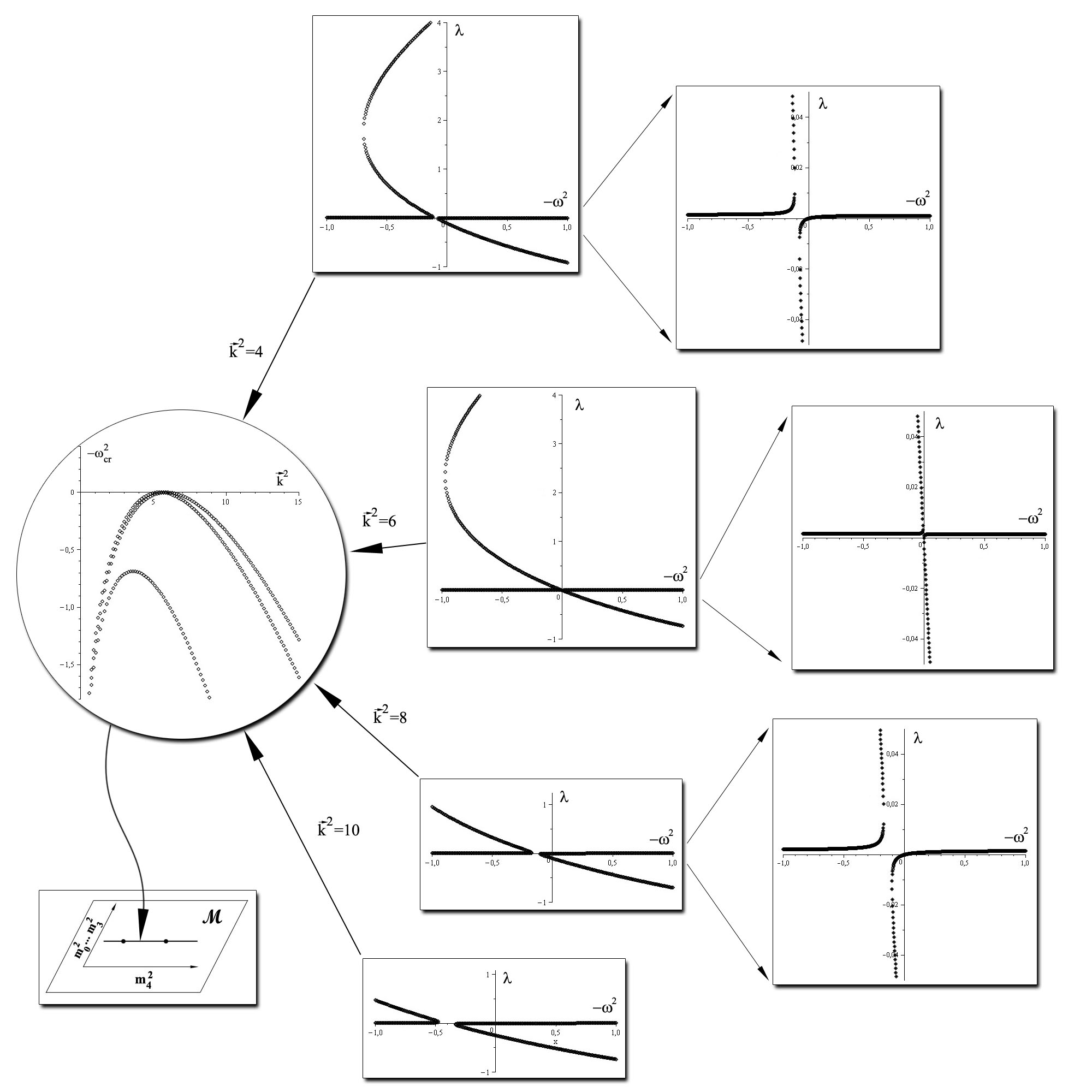}}\caption[t]{\phantom{sjdaskjdhad}}\label{fig4}
\end{center}\end{figure}

At larger value of $m_4^2=3.53553\ldots$ the other
two branches merge as well, and it is quite interesting
to look at the corresponding collection of the eigenvalue plots,
shown enlarged in Fig.\ref{discrC1}.
We see that with the change of $\vec k^2$ the on-shell scalar
converts from normal particle at $\vec k^2<1.623759\ldots$ into
ghost and then back into normal particle at $\vec k^2>11.39337\ldots$.
Note that the \textit{crossing} of branches in discriminant
fiber at $\vec k^2\approx 3.85$ does not cause any reshuffling
in the eigenvalue fiber: this is because this is \textit{crossing}
rather than \textit{merging} of branches.

It deserves mentioning that the slightly-virtual
\textit{instantaneon} can actually be described analytically: at
small values of $\omega^2$ the corresponding eigenvalue is \be
\lambda_{instant} = \frac{2(d-2)\Delta \cdot \omega^2\vec
k^2}{(d-2)(m_3^2-m_2^2)\vec
k^4-\left[(d-3)\Delta+2(d-2)m_2^2m_4^2\right]\vec
k^2+(d-1)m_2^2(m_0^2m_3^2-m_4^4)-m_0^2m_2^4}+ O(\omega^4) \ee
where $\Delta$ is given by (\ref{Delta}). One can
easily check that this simple formula provides a full description of
the bifurcations which we are shown in the above pictures. Thus, the
exactly solvable example confirms the results of
generally applicable discriminant analysis.

\phantom{sdf}

\vspace{1cm}

\newpage

\pagestyle{empty}

\vspace{1cm}

\vspace{1cm}

\vspace{1cm}

\end{document}